\documentclass[aps,prb,twocolumn,groupedaddress, showpacs, longbibiliography, floatfix]{revtex4-1}

\usepackage{amsmath}
\usepackage{amsfonts}
\usepackage{amsthm}
\usepackage{amssymb}

\usepackage{bm}
\usepackage{physics}

\usepackage{graphicx}
\usepackage[caption=false]{subfig}

\usepackage[breaklinks=true,colorlinks,citecolor=blue,linkcolor=red,urlcolor=blue]{hyperref}

\newcommand{\ii}{\mathrm{i}}
\newcommand{\ee}{\mathrm{e}}
\newcommand{\ce}{\varepsilon}
\newcommand{\jp}{J_{\perp}}
\newcommand{\J}{J}
\newcommand{\one}{\openone}
\newcommand{\ash}{\mathrm{arcsinh}}
\newcommand{\ach}{\mathrm{arccosh}}
\newcommand{\fig}{Fig.}
\newcommand{\figs}{Figs.}
\newcommand{\eq}{Eq.}
\newcommand{\eqs}{Eqs.}
\newcommand{\rref}{Ref.}
\newcommand{\rrefs}{Refs.}
\newcommand{\sect}{Section}
\newcommand{\subsect}{Subsection}
\newcommand{\app}{Appendix}

\begin{document}

\title{Motion of an impurity in a two-leg ladder}
\author{Martino Stefanini}
\affiliation{International School for Advanced Studies (SISSA), Via Bonomea 265, 34136 Trieste (Italy) }
\author{Massimo Capone}
\affiliation{International School for Advanced Studies (SISSA), Via Bonomea 265, 34136 Trieste (Italy) }
\author{Alessandro Silva}
\affiliation{International School for Advanced Studies (SISSA), Via Bonomea 265, 34136 Trieste (Italy) }

\date{\today}

\begin{abstract}
 We study the motion of an impurity in a two-leg ladder interacting with two fermionic baths along each leg, a simple model bridging cold atom quantum simulators with an idealised description of the  basic transport processes in a layered heterostructure. Using the linked-cluster expansion we obtain exact analytical results for the single-particle Green's function and find that the long-time behaviour is dominated by an intrinsic orthogonality catastrophe associated to the motion of the impurity in each one-dimensional chain. We explore both the case of two identical legs as well as the case where the legs are characterised by different interaction strengths: in the latter case we observe a subleading correction which can be relevant for intermediate-time transport at an interface between different materials. In all the cases we do not find significant differences between the intra- and inter-leg  Green's functions in the long-time limit.
\end{abstract}

\maketitle

\section{Introduction}

The idea to design devices operating according to the laws of quantum mechanics  holds the promise of a revolution where fundamental quantum phenomena can find a practical, even technological, application\cite{Huelga2013}. 
One of the most natural, yet very promising, perspectives in this context is the possibility to control and enhance transport properties exploiting the synergy between quantum coherence and interactions. An example of transport favoured by coherence is given by molecular biocomplexes where light-harvesting is particularly effective because electronic coherence survives on timescales of the order of a hundred femtoseconds\cite{Scholes2011}.
Extending these ideas from molecules to solids is challenged by the fact that at the high temperatures where devices are meant to work, coherence is expected to be washed out by dephasing associated to interactions with incoherent fluctuations of the lattice, spin and charge degrees of freedom, which are expected to act as a bath.

The fast advances in the engineering of oxide heterostructures based on a few atomic layers\cite{Zubko2011,Hwang2012} can overcome these limitations, as the coherent transport through a very thin layered system can take place over timescales comparable or smaller with respect to the decoherence time\cite{Kropf2019}. 
This calls for theoretical investigations of the fundamental properties of few-layer interacting quantum systems. A realistic many-body calculation for an oxide heterostructure is a very demanding task, which requires advanced numerical methods able to treat strong correlations and the interaction with the excitations of the lattice. Even if the advances of numerical methods able to treat strongly correlated solids make this perspective relatively close, it is of paramount importance to reach some analytical, even if approximate, insight of the basic physical phenomena ruling the coherent transport.

In this work we take the latter perspective, and consider the simplest system realising, at least in principle, a very idealised version of the motion of an excitation in a layered solid, namely an impurity in a two-leg ladder. Each leg of the ladder plays the role of a layer of the heterostructure. The hopping motion from one leg to the other will be investigated as a building block of an interlayer transport. The interaction of the impurity with gapless fermionic degrees of freedom in each leg described as Luttinger liquids is in turn the simplest possible way to describe the effect of interactions on transport.
This choice also allows to connect with well-known properties of impurity problems, including cornerstone phenomena like  the Kondo effect\cite{Kondo} and X-ray edge singularities \cite{Mahan}, and with quantum simulation of impurities in an ultra-cold atom context.

The study of the motion of impurities  has also a very long and successful history, especially focused on the conduction properties of polarons in systems with electron-lattice interactions\cite{Mahan}.
As mentioned above, here we focus on the one-dimensional (1D) version of the problem\cite{PhysRevLett.67.1960,PhysRevLett.68.3638,PhysRevB.46.8858} which allows to introduce interactions in the Fermi system through the Tomonaga-Luttiger Liquid (TLL) theory\cite{Giamarchi,GogolinNersesyanTsvelik}. Moreover, the constrained geometry allows for other simplifications and theoretical techniques\cite{PhysRevB.46.8858} and the existence of exact solutions\cite{McGuire65}.  
    \par The interest in the problem of 1D impurity motion has been witnessing a resurgence in the last two decades\cite{PhysRevB.79.241105,SCHECTER2012639}, thanks to the possibilities offered by ultracold atoms techniques, which have made possible to realise controlled and highly tunable experiments, with various host interactions and impurity types\cite{PhysRevLett.103.150601,PhysRevA.85.023623,Meinert945}. In this perspective, the cold-atom platform also offers a framework to realise idealised versions of the phenomena ruling coherent transport in heterostructures.
    The 1D geometry has proven to be rich of peculiar phenomena, such as pseudo-Bloch oscillations of impurities under an external force\cite{Meinert945} and quantum flutter in the supersonic regime\cite{Mathy:2012aa}. \par From the theoretical point of view, an interesting discovery has been that at low momentum the polaron cannot be simply described as a quasiparticle, contrary to what happens in higher dimensions\cite{Mahan,Rosch1,KSG}, which means that the effect of the bath on the impurity  goes beyond the renormalisation of its properties. The origin of this phenomenon lies in the orthogonality catastrophe (OC)\cite{AndersonOC} caused by the excitation of a large number of low-energy degrees of freedom within the bath.
    \par In this Article we characterise the Green's function of the impurity in the case in which it has access to two 1D chains connected by a tunnelling term. Our main goal is to understand whether the inclusion of the discrete degree of freedom given by the presence of two chains modifies the result of the one-chain problem. We will consider the case of two identical baths and then introduce a difference in the interaction strength, mimicking interfaces between different materials in a heterostructure.
    
    Using a linked-cluster expansion in the interaction strength, we provide detailed calculations and analytic expressions for the Green's function, addressing the difference between the motion within each bath and the motion between the two baths and the connection with the results of a single wire. We anticipate that our results demonstrate that the motion of the impurity remains controlled by the characteristic behaviour of one-dimensional systems, suggesting that a meaningful distinction between inter-wire and intra-wire motion requires larger systems approaching the two-dimensional limit. The system deviates from the behaviour of a single chain when the two wires are different, but this effect is subleading and it can be relevant at intermediate time before the long-time behaviour is recovered. 
    
    The paper is organised as follows: in \sect~\ref{sec:model and results} we introduce the model that we have used to obtain a long-wavelength description of our ladder system, and then we discuss the main results. The following \sect~\ref{sec:LCE} goes deeper into the linked cluster expansion technique and contains more detailed results. \subsect~\ref{subsect:asymptotics} briefly describes how the analytic results on the Green's function have been obtained. \subsect~\ref{subsect:numeric} complements the previous one by illustrating a few examples of the Green's function computed numerically. \subsect~\ref{subsect:spectral function} shows analytic and numeric results on the impurity spectral function, while the following \subsect~\ref{subsect:single bath} displays detailed asymptotic expressions for the single-bath case. Finally, \sect~\ref{sec:conclusions} provides some concluding remarks. Most of the technicalities are discussed in the \app. 
    
\section{Model and main results }\label{sec:model and results}

%
\begin{figure}
     \includegraphics[width=\linewidth]{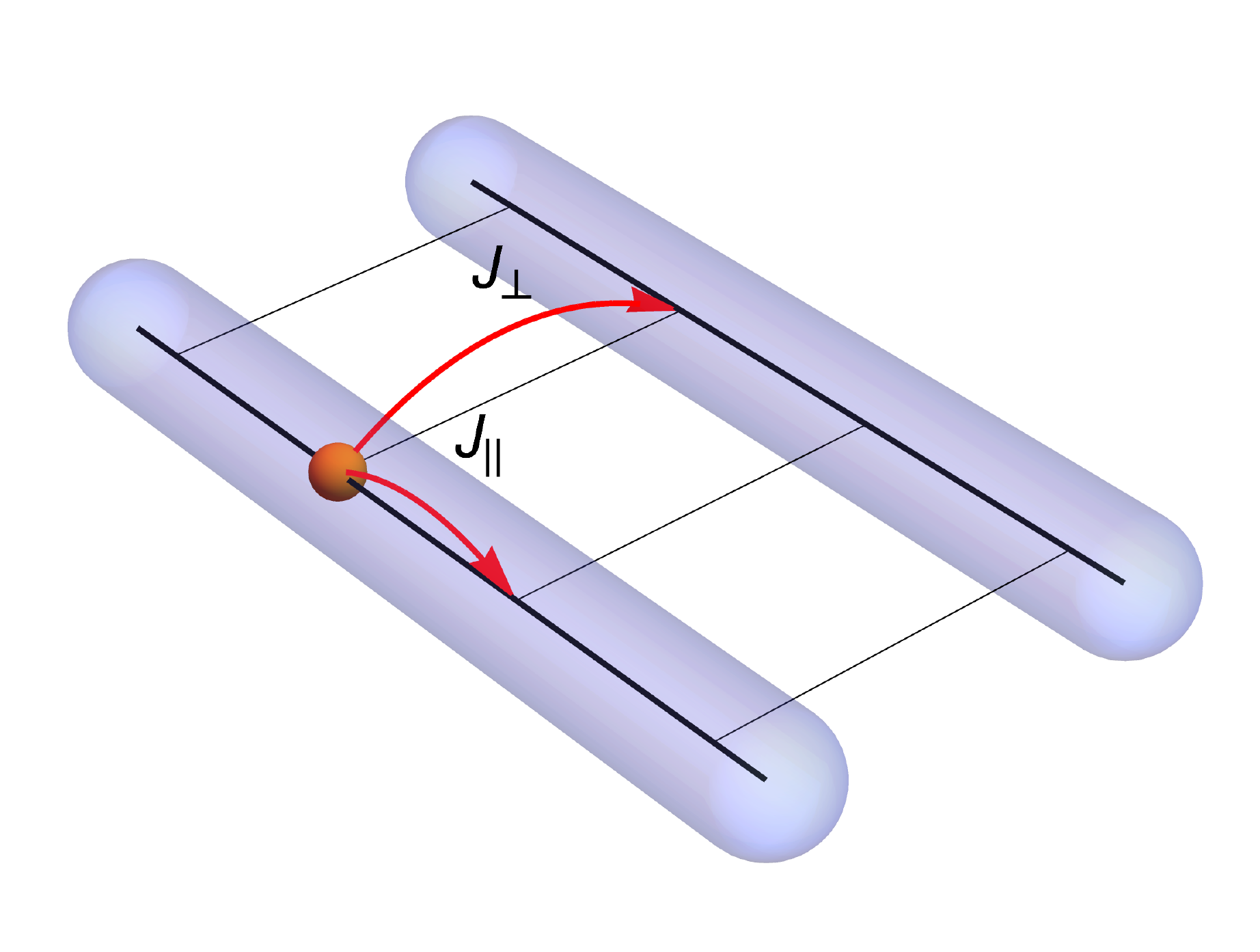}
    \caption{(Color online) The system under analysis is composed of a particle (an impurity) which is able to move through a ladder, whose legs host two independent fermion baths interacting with the impurity.}
    \label{fig:system}
\end{figure}

We are interested in the investigation of the simplest process that can occur in a heterostructure: the dynamics of tunnelling between two subsystems (layers in a three-dimensional heterostructure, or wires in a two-dimensional one) in the presence of strong interactions within each subsystem. In order to focus on the essential features of this process in the following we will consider the dynamics of a spinless impurity which can move both within any of two independent 1D wires (baths) made of interacting fermions (or bosons) or tunnel from one to the other. We make the simplifying assumption that the baths are not directly coupled with each other (our model is thus complementary to the ones analysed in \rrefs~\onlinecite{PhysRevA.100.023614} and \onlinecite{Kamar}, where the baths are coupled). The Hamiltonian has the generic form
\begin{equation}
    \mathcal{H}=\mathcal{H}_{\textup{imp}}+\mathcal{H}_{\textup{bath}}+\mathcal{H}_{\textup{c}}~,
\end{equation}
where the three terms describe the free impurity, the  baths and the impurity-bath coupling, respectively. The impurity motion can be thought of as happening on a ladder (see \fig~\ref{fig:system} ) where $J_{\parallel}$ and $J_{\perp}$ are the hopping matrix elements along and between the chains respectively. In the present article we are interested in the continuum description of such a system, hence the conclusions we will reach hold irrespective of the presence of a lattice. In the long-wavelength limit the impurity dispersion is  parabolic\footnote{We use units such that $\hbar=1$}:
\begin{gather}
    \mathcal{H}_{\textup{imp}}=\sum_{\sigma}{}\int{\dd{x}[d^{\dag}_{\sigma}(x)E(-\ii\dv*{}{x})d_{\sigma}(x)}+\nonumber\\-\jp d^{\dag}_{\bar{\sigma}}(x)d_{\sigma}(x)]~,\\
    \intertext{with}
    E(p)=\frac{p^{2}}{2M}~.\nonumber
\end{gather}
with $M=1/(2J_{\parallel}a^2)$, $a$ being the lattice spacing.
The pseudo-spin index $\sigma=\pm1$ labels the two chains ($\bar{\sigma}=-\sigma$), and $d_\sigma(x)$ is a fermionic field (but all subsequent results equally apply to a bosonic impurity, because statistics is irrelevant for a single particle). We assume that the system is confined on a segment of length $L$, obeying periodic boundary conditions. At long wavelength, the baths are described by two independent Tomonaga-Luttinger Liquid (TLL) Hamiltonians \cite{Giamarchi} 
\begin{align*}
    \mathcal{H}_{\textup{bath}}&=\sum_{\sigma}{u_{\sigma}\int{\frac{\dd{x}}{2\pi}}\bigg[K_{\sigma}\bigg(\dv{\theta_{\sigma}}{x}(x)\bigg)^{2}+\frac{1}{K_\sigma}\bigg(\dv{\phi_{\sigma}}{x}(x)\bigg)^{2}\bigg]}~,
\end{align*}
where $u_\sigma$ is the sound speed in bath $\sigma$, while the Luttinger parameter $K_\sigma$ measures the interaction of the particles constituting the bath. In particular, for a fermionic bath, $K_\sigma<1$ ($K_\sigma>1$) for repulsive (attractive) interactions. Finally, the bath-impurity coupling is chosen as the minimal one, a simple density-density interaction
\begin{equation}
    \mathcal{H}_{\textup{c}}=\sum_{\sigma}{g_{\sigma}\int{\dd{x}d_{\sigma}^{\dag}(x)d_{\sigma}(x)\rho_{\sigma}(x)}}~,
\end{equation}
with $\rho_\sigma(x)$ being the particle density of the bath $\sigma$. At the level of our treatment (long wavelength, weak coupling), this interaction can be either repulsive or  attractive (i.e. $g_\sigma>0$ or $<0$), as everything will only depend on $g_\sigma^2$. We will generally take $g_\sigma>0$ for concreteness. 
\par Bosonization\cite{Giamarchi} provides a link between density fluctuations and the boson field $\phi$:
\begin{gather}\label{eq:bosonized density}
\rho_{\sigma}(x)=\rho_{\sigma}^{(0)}-\frac{1}{\pi}{\dv{}{x}\phi_{\sigma}}(x)+\nonumber\\
+\rho_{\sigma}^{(0)}\sum_{n\neq0}{\ee^{2n\ii(\pi\rho_{\sigma}^{(0)}x-\phi_{\sigma}(x))}}~,
\end{gather}
$\rho^{(0)}_\sigma$ being the average density of the bath $\sigma$. As the third term describes oscillations with large wavenumbers $2n\pi\rho^{(0)}_\sigma$, we keep only the first two. The discarded terms are important only in case we want to describe fast impurities, or effects like the pseudo-Bloch oscillations. The constant term can be adsorbed in $\phi_\sigma$ by a canonical transformation, hence the final expression of the coupling Hamiltonian is
\begin{equation}
     \mathcal{H}_{\textup{c}}=-\sum_{\sigma}{\frac{g_{\sigma}}{\pi}\int{\dd{x}d_{\sigma}^{\dag}(x)d_{\sigma}(x)\dv{}{x}\phi_{\sigma}(x)}}~.
\end{equation}

Following \rref~\onlinecite{Giamarchi}, we express the $\phi_\sigma$ field in terms of phonon modes\footnote{We are implicitly dropping terms of $\order{1/L}$, that give vanishing contributions in the thermodynamic limit.} $b_{q\sigma}$:
\begin{equation}
    \phi_\sigma(x)=-\ii\pi\frac{K_\sigma^{1/2}}{L^{1/2}}\sum_{p\neq0}{\frac{V(p)}{p}\ee^{-\ii p x}(b^\dag_{p\sigma}+b_{-p\sigma})}
\end{equation}
and we obtain the Hamiltonian in momentum space:
\begin{subequations}
\begin{align}
\mathcal{H}_{\textup{imp}}&=\sum_{p\sigma}{\qty(E(p)d^{\dag}_{p\sigma}d_{p\sigma}-\jp d_{p\bar{\sigma}}^{\dag}d_{p\sigma})}~,\\
\mathcal{H}_{\textup{bath}}&=\sum_{p\neq0,\sigma}{u_{\sigma}\abs{p}b_{p\sigma}^{\dag}b_{p\sigma}}~,\\
\mathcal{H}_{\textup{c}}&=\sum_\sigma{\frac{g_\sigma K_{\sigma}^{1/2}}{L^{1/2}}\sum_{p\neq0}{V(p)N_{\sigma}(p)(b_{p\sigma}^{\dag}+b_{-p\sigma})}}~,
\end{align}
\end{subequations}
In the above formulae, 
\[
V(p)=\qty(\frac{\abs{p}}{2\pi})^{1/2}\ee^{-\alpha p/2}~,
\]
and $\alpha$ is a small length providing an ultraviolet momentum cutoff (it can be identified with the underlying lattice constant), and $N_{\sigma}(p)=\sum_{k}{d^{\dag}_{k-p,\sigma}d_{k\sigma}}$ is the Fourier transform of the impurity density.
\par As one can see, the role of Luttinger parameters $K_\sigma$ is to rescale the bare interaction to 
\begin{equation}
  \Tilde{g}_\sigma=g_\sigma K_\sigma^{1/2}~.
\end{equation}
This implies that baths with attractive interactions ($K_\sigma>1$) are coupled more strongly to the impurity than baths whose particles repel each other ($K_\sigma<1$).

\par We compute the impurity Green's function.
\begin{equation}\label{eq:def G}
G_{\sigma\prime\sigma}(p^\prime,p;t)=-\ii\ev{\mathcal{T}d_{p^\prime\sigma^\prime}(t)d_{p\sigma}^{\dag}}{\Omega}~,
\end{equation}
where $d_{p\sigma}(t)=\ee^{\ii\mathcal{H}t}d_{p\sigma}\ee^{-\ii\mathcal{H}t}$, $\mathcal{T}$ is the time-ordering symbol and the vector $\ket{\Omega}=\ket{0}_{d}\ket{\omega}_{b}$ is the product of the impurity vacuum $\ket{0}_{d}$ and the interacting bath ground state $\ket{\omega}_{b}$. Here $ G_{\sigma\prime\sigma}(p^\prime,p;t)$ describes indeed the process in which we prepare the baths in their ground state $\ket{\omega}_b=\prod_{\sigma}{\ket{\omega_\sigma}_b}$ and, at time $t=0$, we inject an impurity with a definite momentum $p$ in the chain $\sigma$, probing the subsequent evolution by focusing on the amplitude for elastic scattering, i.e. the overlap of the state vector with every possible one-impurity state with no phonons.
We notice that in ultra-cold atomic experiments the  return amplitude, $\ii G_{\sigma\sigma}(p,p;t)$, is a measurable quantity, both in real time (through interferometry) and in frequency (it gives the absorption cross-section in radio-frequency spectroscopy)\cite{PhysRevX.2.041020,PhysRevA.89.053617} 
\par In the following we will present a perturbative scheme to compute $G_{\sigma^\prime\sigma}(p^\prime,p;t)$. We notice that $\ket{\Omega}$ is the ground state of the full $\mathcal{H}$ in the Hilbert space sector without impurities, i.e., $\mathcal{H}\ket{\Omega}=0~$, 
and it coincides with the \emph{noninteracting} groundstate within the same sector of the Hilbert space. This property, combined with the fact that the time evolution conserves the number of impurities allows us to use the Gell-Mann\&Low theorem\cite{Mahan} to compute \eq~\eqref{eq:def G} within the standard zero-temperature perturbation theory, despite the fact that the process we are investigating is actually out of equilibrium.
The perturbative series is built using $\mathcal{H}_\textup{imp}+\mathcal{H}_\textup{bath}$ as  unperturbed Hamiltonian (hence, the results will be nonperturbative in $\jp$), and expanding in the coupling $\tilde{g}_\sigma$\footnote{It can be easily seen that $g_\sigma$ has the dimension of a speed, so that the dimensionless coupling constant is actually $\tilde{g}_\sigma/u_\sigma$. Compare with the expression for $\beta(p)$ (\eq~\eqref{eq:beta}).}. Because of translational invariance it is convenient to work in momentum space, and it is straightforward to see that $G_{\sigma\prime\sigma}(p^\prime,p;t)=\delta_{p^\prime,p}G_{\sigma\prime\sigma}(p,t)$ is actually diagonal in $p$. 
The detailed calculation of the Green's function is rather involved, thus we provide here the main results of our analysis.
\par As a benchmark let us start our analysis with the noninteracting system. In the absence of baths the Hamiltonian of the impurity can be easily diagonalised in terms of symmetric and anti-symmetric modes $d_{p,e/o}=(d_{p,+}\pm d_{p,-})/\sqrt{2}$ with dispersion
\begin{equation}
    \lambda_{e,o}(p)=E(p)\mp\jp~.
\end{equation}
The resulting Green's function is therefore the sum of two quasi-particle contributions associated to the two sub-bands 
\begin{equation}\label{eq:G0}
    (G_0)_{\sigma\sigma^\prime}(p,t)=-\tfrac{\ii }{2}\theta(t)(\ee^{-\ii\lambda_e(p)t}+\sigma\sigma^\prime \ee^{-\ii\lambda_o(p)t}).
\end{equation}
In the presence of interactions within the baths we expect two main physical effects: an orthogonality catastrophe associated to the bath response to the injection of the impurity\cite{KSG,Giamarchi,GogolinNersesyanTsvelik,PhysRevLett.75.1988,PhysRevB.79.241105} and dissipation due to the excitation of phonons in the baths. These effects can be consistently captured by a standard Linked-Cluster expansion\cite{Mahan,KSG}\footnote{In Ref. \onlinecite{Kamar} an analogous calculation is performed in a generalisation of the model we are studying, with the complication that also the fermions constituting the baths can hop between the legs of the ladder.}. The outline of the derivation is presented in the following \subsect. We find that, at subsonic momenta $\abs{p}<M\min_\sigma\{u_\sigma\}$ and long times $t\gg\jp^{-1},(Mu_\sigma^2)^{-1}$, the Green's function has the asymptotic expression
\begin{align}\label{eq: G asymp}
    G_{\sigma^\prime\sigma}(p,t)\sim-\tfrac{\ii}{2}\qty(\tfrac{t_0}{t})^{\beta(p)}\Big(Z_e(p,t_0)\ee^{-\ii\Tilde{\lambda}_e(p)t}+\nonumber\\
  +\sigma^\prime\sigma Z_o(p,t_0)\ee^{-2\gamma(p)t-\ii\Tilde{\lambda}_o(p)t}\Big)\big(1+\order{\tfrac{1}{t}}\big)~
\end{align}
where $\Tilde{\lambda}_{e,o}(p)$ are the renormalised bands, $t_0$ is an arbitrary time scale and the expressions for all coefficients can be found in the \app. The above Green's function shows the effects mentioned before. The second factor is a renormalised version of the noninteracting Green's function. This quasiparticle behaviour is spoiled by the first factor, whose power-law decay in time is the typical manifestation of the OC\cite{Mahan}. 
\par Being an excited state, the anti-symmetric band becomes unstable, decaying exponentially with a rate $2\gamma(p)$ given by
\begin{equation}\label{eq:gamma}
2\gamma(p)=\frac{M}{8\pi}\sum_{\sigma,s=\pm 1}{\Tilde{g}_\sigma^2\Bigg(1-\frac{1}{\sqrt{1+2\jp/k_{s\sigma}(p)}}\Bigg)}~,\\
\end{equation}
where $k_{s\sigma}(p)=(Mu_{\sigma}+sp)^2/2M$. Notice that for $J_{\perp}=0$ this expression vanishes: this is consistent with the fact that in the absence of tunnelling there cannot be emission of phonons by an impurity with subsonic speed since momentum and energy conservation cannot be simultaneously satisfied. This simple fact hinders any decay in the symmetric, low energy band, while emission of phonons with a decay from the anti-symmetric one is possible, hence the finite relaxation rate.  
\par Interestingly, the asymmetry observed in the physics above is not present in the power-law decay associated to the OC occurring as the 1D bath rearranges in response to the injection of the impurity.
The Green's function is characterised by a \emph{single} OC exponent $\beta(p)$:
\begin{equation}\label{eq:beta}
    \beta(p)=\frac{1}{8\pi^2}\sum_{\sigma}{\frac{g_\sigma^2K_\sigma}{u_\sigma^2}\frac{1+(p/Mu_\sigma)^2}{(1-(p/Mu_\sigma)^2)^2}}~,
\end{equation}
which is the same for all components. A comparison with \rref~\onlinecite{KSG} shows that it is proportional to the sum of the analogous single-bath exponents $\beta^\textup{sb}_\sigma(p)$ (see also \subsect~\ref{subsect:single bath}),
\begin{equation}\label{eq: beta vs sb}
    \beta(p)=\tfrac{1}{4}(\beta^\textup{sb}_+(p)+\beta^\textup{sb}_-(p))~.
\end{equation}
Thus, this exponent is half the average of the  two single-chain ones and, interestingly, it does not depend on the inter-chain hopping $\jp$. We want to stress that the above results are non perturbative in $\jp$. We are lead to the conclusion that the mere addition of the bath degree of freedom is able to weaken the orthogonality catastrophe, but not to destroy it. To be more explicit, if the baths have identical properties then $\beta(p)=\beta^{\textup{sb}}/2$. 
\par The fact that there is only a single OC exponent characterising both baths is related to the fact that this phenomenon is observed in the limit $\jp t\gg1$, that is, when the impurity has had enough time to repeatedly interact with each bath. In fact, the numeric evaluation of the Green's function shows that for short times $\jp t\ll1$ there is a ''dimensional crossover''. At $t=0$, $G_{\sigma\sigma}$ evolves close to the single-bath $G_\sigma^{\textup{sb}}$, rapidly establishing its characteristic power-law $\beta^\textup{sb}_\sigma$. Then, the impurity starts to populate the other bath and the Green's function acquires the two-bath shape.
\par We are inclined to think that the reason for $\beta(p)$ being less than the average of the single-bath exponents is precisely related to the fact that at weak coupling the impurity is able to spread across both baths. As a hand-waving argument, we may think that each bath effectively sees only ''half'' of the impurity, so that the actual couplings become $\tilde{g}_\sigma/2$. In this picture, \eq~\eqref{eq: beta vs sb} would only state that $\beta(p)$ is the sum of the single-bath exponents computed with $\tilde{g}_\sigma/2$ instead of $\tilde{g}_\sigma$, because all of these quantities are proportional to $\tilde{g}_\sigma^2$. This line of reasoning is supported by the extension to the case of $N$ baths, in which it would yield $\beta^N(p)=\sum_i{\beta^{\textup{sb}}_i}/N^2$. In the limit $N\to\infty$ the exponent would vanish. In such a limit the impurity motion would become effectively two-dimensional, while the baths would still be bosonic, and so we would not get any OC.
\par Let us also comment on the fact that the noticeable divergence of $\beta(p)$ close to the threshold for phonon emission $p=Mu_\sigma$ is in a range of momenta were we do not expect the second-order LCE and maybe even the long-wavelength approximation to be sufficient. 


\par From the expression \eq~\eqref{eq: G asymp} we can also calculate the spectral function $\hat{A}(p,\omega)\equiv-2\Im\hat{G}(p,\omega)$, finding that for $\beta(p)<1$ (that is, for small coupling and momentum) it has a diverging threshold at $\omega=\Tilde{\lambda}_e(p)$,
\begin{equation}
    A_{\sigma\sigma}(p,\omega)\sim\theta(\omega-\Tilde{\lambda}_e(p))\frac{1}{(\omega-\Tilde{\lambda}_e(p))^{1-\beta(p)}}~.
\end{equation}
This is the how the power-law decay in time, hence the OC,  manifests itself in the frequency domain. At higher frequency, the spectral function has a non-Lorentzian peak at the excited mode $\Tilde{\lambda}_o(p)$. More details can be found in \subsect~\ref{subsect:spectral function}.

\section{The Linked-Cluster Expansion}\label{sec:LCE}
In single-particle problems, and especially in impurity models, the so-called Linked-Cluster Expansion\cite{Mahan,KSG} has been historically successful in providing very good approximations for the Green's function, even at intermediate coupling.
\par  At the lowest nontrivial order, the LCE amounts to 
\begin{equation}
    \hat{G}(p,t)=\hat{G}_0(p,t)\cdot\ee^{\hat{F}_2(p,t)}~,
\end{equation}
where symbols with a hat will represent matrices in chain index space $(\sigma,\sigma^\prime)$, and
\begin{equation}\label{eq: G0}
    \hat{G}_{0}(p,t)=-\ii\theta(t)\mqty(\cos{\jp t} & \ii\sin{\jp t}\\ \ii\sin{\jp t} & \cos{\jp t})\ee^{-\ii E_{p} t}
\end{equation}
is the noninteracting impurity Green's function, $\cdot$ is the matrix product and $\hat{F}_2(p,t)$ is defined by
\begin{equation}\label{eq: def F2}
    \hat{G}_0(p,t)\cdot\hat{F}_2(p,t)=(\hat{G}_0\ast\hat{\Sigma}^{(2)}\ast\hat{G}_0)(p,t)~.
\end{equation}
In the above equation, $\ast$ represents a time convolution, while $\hat{\Sigma}^{(2)}(p,t)$ is the second-order self-energy\footnote{Within our model there is no tadpole contribution, because $\hat{G_0}(p,0^-)=0$ and because there are no zero-momentum phonons}:
\begin{gather}\label{eq: structure SE}
     \hat{\Sigma}^{(2)}(p,t)=\mqty(\Sigma_{+}(p,t) & 0 \\ 0 & \Sigma_{-}(p,t)),\\
     \intertext{where}
     \begin{split}\label{eq: expression SE}         \Sigma_\sigma(p,t)=\ii g_\sigma^2 K_\sigma\frac{1}{L}\sum_{q\neq0}{V^2(q)(G_0)_{\sigma\sigma}(p-q,t)D_\sigma^0(q,t)}=\\
    =-\ii \Tilde{g}_\sigma^2\theta(t)\cos{\jp t}\,\int{\tfrac{\dd{q}}{2\pi}V^2(q)\, \ee^{-\ii(E(p-q)+u_\sigma\abs{q})t}}~,
     \end{split}
\end{gather}
with
\begin{multline}
    D_{\sigma}^{0}(p,t)\equiv\\-\ii\ev{\mathcal{T}(b^\dag_{p\sigma}(t)+b_{-p\sigma}(t))(b^\dag_{-p\sigma}+b_{p\sigma})}{\Omega}_{\mathcal{H}_\textup{bath}}=\\
    =-\ii\theta(t)\ee^{-\ii u_{\sigma}\abs{p}t}-\ii\theta(-t)\ee^{\ii u_{\sigma}\abs{p}t}~.
\end{multline}
\par As one can see, we have chosen to work directly in the bath basis, where the physics is more transparent. In this basis, only the free impurity propagation is able to change the bath index (as $\hat{G}_0(p,t)$ is not diagonal, \eq\eqref{eq: G0}), whereas impurity-bath interactions conserve $\sigma$. 
\par In order to compute $\hat{F}_2$ from \eq~\eqref{eq: def F2}, it is convenient to first switch to the basis in which $\hat{G}_0(p,t)$ is diagonal. Most importantly, it is advantageous to perform the momentum integration in the self-energy Eq.\eqref{eq: expression SE} after the time convolution. The result of this calculation is best expressed by decomposing $\hat{F}_2$ in the matrix Pauli matrices basis $\hat{\bm{\sigma}}\equiv(\sigma_1,\sigma_2,\sigma_3)$, along with the $2\times2$ identity matrix $\one$
\begin{equation}\label{eq:full F2}
    \hat{F}_2(p,t)=A(p,t)\one +(B(p,t),C(p,t),D(p,t))\cdot\hat{\bm{\sigma}}~,
\end{equation}
where $A,\,B,\,C,\,D$ are complex functions which are defined in \app A. The exponential is now easily computed (omitting the $(p,t)$ arguments):
\begin{equation}\label{eq:expF2}
    \ee^{\hat{F}_2(p,t)}=\ee^{A}\Big(\cosh{\lambda}\,\one+\frac{\sinh{\lambda}}{\lambda}(B,C,D)\cdot\hat{\bm{\sigma}}\Big)~,
\end{equation}
where
\begin{equation}\label{eq:lambda}
    \lambda(p,t)\equiv\sqrt{B^2(p,t)+C^2(p,t)+D^2(p,t)}
\end{equation}
(any of the two complex roots can be chosen).  
\par The meaning and physical role of the above functions will become clear in the next \sect, but for now we point out that the the $C(p,t)$ and $D(p,t)$ functions are nonzero only if there is an asymmetry between the baths (in any of the $u_\sigma,\, K_\sigma$ or in $g_\sigma$). The other $A(p,t)$ and $B(p,t)$ functions are less sensitive to asymmetries, and will be shown to encapsulate most of the physics of the problem.
\par Putting together \eq~\eqref{eq:expF2} and \eq~\eqref{eq: G0} we obtain the full expression for the Green's function:
\begin{subequations}\label{eq: full G}
\begin{align}
    G_{\sigma\sigma}(p,t)&=-\ii\theta(t)\ee^{-\ii E(p)t+A(p,t)}\times\nonumber\\
    [\cosh{\lambda}&\cos(\jp t)+\ii\tfrac{\sinh{\lambda}}{\lambda}B\sin(\jp t)+\nonumber\\
    +&\sigma\tfrac{\sinh{\lambda}}{\lambda}(D\cos(\jp t)-C\sin(\jp t))]~,\\
    G_{\sigma\bar{\sigma}}(p,t)&=-\ii\theta(t)\ee^{-\ii E(p)t+A(p,t)}\times\nonumber\\
    &[\ii\cosh{\lambda}\sin(\jp t)+\tfrac{\sinh{\lambda}}{\lambda}B\cos(\jp t)]~,
\end{align}
\end{subequations}
This formula is rather general and thus not very enlightening, but it 
shows that the off-diagonal elements of $\hat{G}(p,t)$ are equal, just as in the noninteracting case. This is rooted in the fact is that we used the self-energy computed at second order in perturbation theory, which is diagonal in bath index space (see \eq~\eqref{eq: structure SE}). Going to the next perturbative order (the fourth) allows for the inclusions of vertex corrections, which generally provide\footnote{However, this may not happen if $\Tilde{g}_+\neq \Tilde{g}_-$ but the sound speeds $u_\sigma$ are the same.} $\hat{\Sigma}$ with off-diagonal elements.
\par In principle, given the expressions of $A(p,t)-D(p,t)$ in \app, we have all the ingredients to compute the second-order LCE Green's function. In general, $A,\,B,\,C,\,D$ are defined by integrals which have to be computed numerically. The results of such computations will be described ins \sect~\ref{subsect:numeric}. However, we can obtain rather accurate asymptotic approximations at long times $t\gg (Mu_\sigma^2)^{-1},\, \jp^{-1}$, as we show in the next \sect. In this way we will understand the physical content of the Green's function that we obtained. 
\subsection{Asymptotic behaviour of the Green's function}
\label{subsect:asymptotics}

In this \sect we build on our exact expression for $\hat{G}(p,t)$ to gain some physical insight. We will first show that its structure can be simplified and rationalised in the long-time limit, then we will display a few examples of the full numerical computation. 
\par As stated above, we will present only results concerning subsonic impurities, for which $\abs{p}$ is smaller than both $Mu_+$ and $Mu_-$. We are interested in this regime for two reasons: First, we do not expect our model, nor the second-order LCE to be accurate when $p$ is around $Mu_\sigma$, as the system becomes effectively strongly interacting. Second, as the threshold is exceeded, real phonons begin to be emitted in all bands and the Green's function vanishes exponentially in time thus loosing any usefulness. In fact, we have extended our calculations to this regime, finding that this happens in less than a period $2\pi/\jp$. 
%
%
\par We can recast the $\hat{F}_2(p,t)$ function obtained in LCE in the following asymptotic form:
\begin{equation*}
    \hat{F}_2(p,t)\sim \hat{X}(p)t+\hat{Y}(p)+\hat{\phi}_\textup{nl}(p,t)+\order{\tfrac{1}{t}}
\end{equation*}
for $t\to\infty$. On the right-hand side of this equation, the first two terms describe quasiparticle physics: the $\hat{X}(p)$ matrix will renormalize the single-particle properties (i.e. the mass and inter-bath hopping $\jp$) and possibly give a finite lifetime to the momentum state. The $\hat{Y}(p)$ matrix will instead quantify the (matricial) quasiparticle residue $\hat{Z}=\exp\hat{Y}$. The third term stands for any possible subleading nonlinear function of time. If any of such terms is present, it causes the Green's function to depart from the quasiparticle picture. In two and three dimensions $\hat{F}_2$ lacks any nonlinearity, and the polaron behaves as a quasiparticle\cite{Mahan,Rosch1}. In one dimension and with one bath, it was shown\cite{KSG,Rosch1} that there is a logarithmic term, which causes the Green's function to acquire a power-law decay at long times. This is related to the orthogonality catastrophe\cite{AndersonOC,Mahan,Giamarchi, GogolinNersesyanTsvelik,PhysRevB.79.241105,PhysRevX.2.041020}. 
\par In the present problem the asymptotic expansion of $F_2$ can be obtained from that of $A,\,B,\,C,\,D$ using Eq.(\ref{eq:expF2}).
As shown in \app~\ref{app: calculations}, the four functions have the following asymptotic form:
\begin{subequations}\label{eq: asympt ABCD}
\begin{align}
    A(p,t)&\sim -\gamma(p) t-\ii\Delta E(p)t+c_A(p,t_0)+\nonumber\\
    &-\beta(p)\ln{\tfrac{t}{t_0}}+\order{\tfrac{1}{t}}\\
    B(p,t)&\sim\gamma(p) t+\ii\Delta \J(p)t +c_B(p)+\order{\tfrac{1}{t}}\\
    C(p,t)&\sim\ii\frac{1-\cos{2\jp t}}{\jp}c_H^{(0)}+\nonumber\\
    &+\frac{\sin{\jp t}}{\jp}\big(c_H^{(-)}\ee^{\ii\jp t}-c_H^{(+)}\ee^{-\ii\jp t}\big)+\order{\tfrac{1}{t}}\\
    D(p,t)&\sim -\ii\frac{\sin{2\jp t}}{\jp}c_H^{(0)}+\nonumber\\
    &+\frac{\cos{\jp t}}{\jp}\big(c_H^{(+)}\ee^{-\ii\jp t}-c_H^{(-)}\ee^{\ii\jp t}\big)+\order{\tfrac{1}{t}}~,
\end{align}
\end{subequations}
where $t_0$ is an arbitrary time scale, and the expressions for all the coefficients $c_{A,B},\,c^{\pm, 0}_H$ and $\Delta J,\,\Delta E$ can be found in the \app. We found numerically that for momenta close to zero the approximations given above provide an excellent approximation to the Green's function for $\jp t\agt0.1$, i.e. for almost every relevant time. This accuracy is lost only for significantly high momenta $\abs{p}\agt0.8Mu_\sigma$, but even in this case \eqs~\eqref{eq: asympt ABCD} become reliable if $t$ is large enough.  
\par According to the above general considerations, the leading behaviour of the $A(p,t)$ and $B(p,t)$ functions is linear in time (however, notice the logarithmic term in the former), whereas the $C(p,t)$ and $D(p,t)$ functions are purely oscillating. Therefore, at sufficiently large times the first two are much larger than the last two, and we can approximate \eq~\eqref{eq:lambda}
\begin{equation*}
    \lambda\equiv\sqrt{B^2+C^2+D^2}=B+\order{\tfrac{1}{t}}
\end{equation*}
obtaining the expressions 
\begin{subequations}\label{eq: G asymp1}
\begin{align}
    G_{\sigma\sigma}(p,t)\sim-\ii\,\ee^{c_A(p)-\gamma(p)t-\ii\Tilde{E}(p)t}(t/t_0)^{-\beta(p)}\times\nonumber\\
    \cosh(c_B(p)+\gamma(p)t+\ii\Tilde{J}_\perp(p) t)\big(1+\order{\tfrac{1}{t}}\big)~,\\
    G_{\sigma\bar{\sigma}}(p,t)\sim-\ii\,\ee^{c_A(p)-\gamma(p)t-\ii\Tilde{E}(p)t}(t/t_0)^{-\beta(p)}\times\nonumber\\
    \sinh(c_B(p)+\gamma(p)t+\ii\Tilde{J}_\perp(p) t)\big(1+\order{\tfrac{1}{t}}\big)
\end{align}
\end{subequations}
where 
\begin{subequations}\label{eq: renormalized E J}
\begin{align}
    \Tilde{E}(p)&\equiv E(p)+\Delta E(p)~,\\
    \Tilde{J}_\perp(p)&\equiv \jp+\Delta \jp(p)~,
\end{align}
\end{subequations}
are the single-particle properties modified by interactions. Notice the similarity with the noninteracting expression \eq~\eqref{eq: G0}. Of course, the main difference is the already mentioned power-law decay, whose exponent has been already reported in \eq~\eqref{eq:beta}.
\par A surprising property of \eq~\eqref{eq: G asymp1} is that is that at late times $G_{++}\sim G_{--}$ ( $G_{+-}=G_{-+}$ is guaranteed), i.e. the effect of any difference in the baths is apparently washed out by $J_{\perp}$. In fact, this difference is hidden in part of the subleading $\order{1/t}$ terms which have been omitted, proportional to the $C(p,t)$ and $D(p,t)$ functions.
\par The difference in the baths' parameters manifests itself in the functions $C(p,t)$ and $D(p,t)$ which oscillate in time, combining frequencies that are small multiples of the fundamental $\jp$. As it will be clear in the next \sect, their contribution is essentially a high-frequency "noise" at small and intermediate times, which is gradually erased. 
\par The expression \eq~\eqref{eq: G asymp} is recovered by simply expanding the hyperbolic functions into exponentials. We recall here its structure:
\begin{multline}\label{eq: G asymp2}
    G_{\sigma\sigma^\prime}(p,t)\sim-\tfrac{\ii}{2}\qty(\tfrac{t_0}{t})^{\beta(p)}
    \Big(Z_e(p,t_0)\ee^{-\ii\Tilde{\lambda}_e(p)t}+\nonumber\\
    \sigma\sigma^\prime Z_o(p,t_0)\ee^{-2\gamma(p)t-\ii\Tilde{\lambda}_o(p)t}\Big)\big(1+\order{\tfrac{1}{t}}\big)~,
\end{multline}
where we introduced the complex "quasiparticle weights" $Z_{e,o}(p,t_0)\equiv\exp(c_A(p,t_0)\pm c_B(p))$ and the renormalised bands 
\begin{equation}
    \tilde{\lambda}_{e,o}=\tilde{E}(p)\mp\tilde{J}_\perp(p)~.
\end{equation}
The contribution from the symmetric mode (which is the ground state of the system at total momentum $p$) decays slowly, only as a power-law\footnote{Notice that for momenta $p\ll Mu_\sigma$ and small coupling $\Tilde{g}_\sigma/u_\sigma<1$, $\beta(p)$ is a small number of the order of $10^{-2}$.}, whereas the other mode, being of higher energy, is further suppressed by an exponential decay with a characteristic time $(2\gamma(p))^{-1}$.
\par Up to now, we did not discuss explicitly the cutoff dependence of our results. Important measurable quantities, like $\beta(p)$ and $\gamma(p)$, are evidently independent of the high-energy properties of the baths. This does not hold for the whole Green's function. Yet $\Lambda$ enters only in two terms, through $\ln{\Lambda}$ in both cases. The first is an irrelevant overall energy shift $\Delta E(p)$. The second dependence is within the $C(p,t)$ and $D(p,t)$ functions and their asymptotic coefficients $c_H^{0,\pm}$. Hence, while in the case of equal baths $G_{\sigma^\prime,\sigma}(p,t)$ is well-defined except for an overall phase factor, when the baths are different the Green's function is cutoff-dependent but for very long times. 
\par We end this section by noticing an interesting relation, valid for $\jp\to0$:
\begin{equation}
\gamma(p)=\pi\beta(p)\jp+\order{\jp^2}~.    
\end{equation}
What makes it noteworthy is that $\gamma(p)$ is ''easy'' to compute, as it can be obtained by a straightforward Fermi Golden Rule calculation, while $\beta(p)$ can be obtained only by summing infinite subdiagrams in perturbation theory (which is exactly what the LCE does).
\subsection{Numerical results}\label{subsect:numeric}
\begin{figure}
    \centering
    \subfloat[][
    \label{fig:Gsym p}]{\includegraphics[width=.9\linewidth]{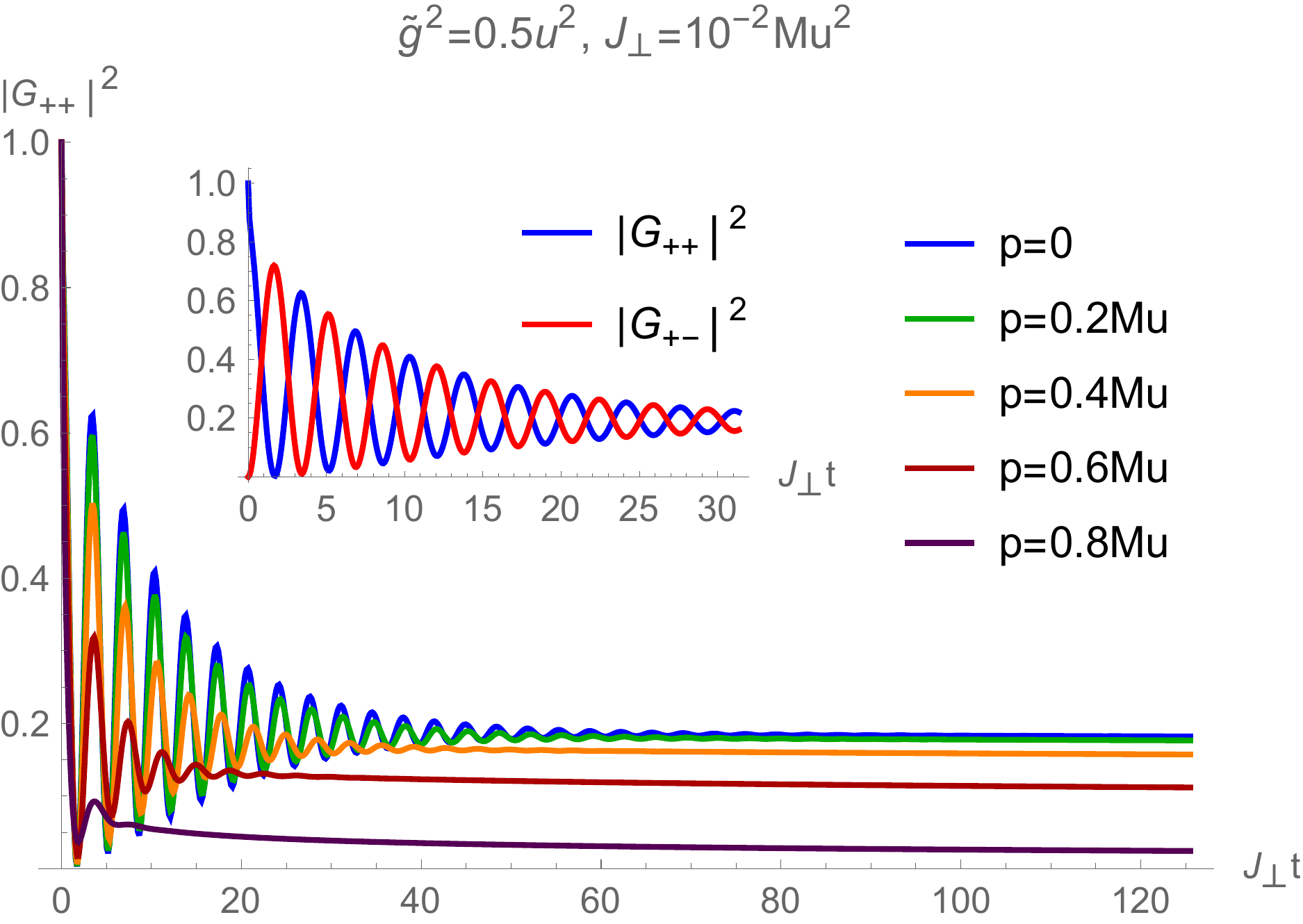}}\\
    \subfloat[][
    \label{fig:Gsym J}]{\includegraphics[width=.9\linewidth]{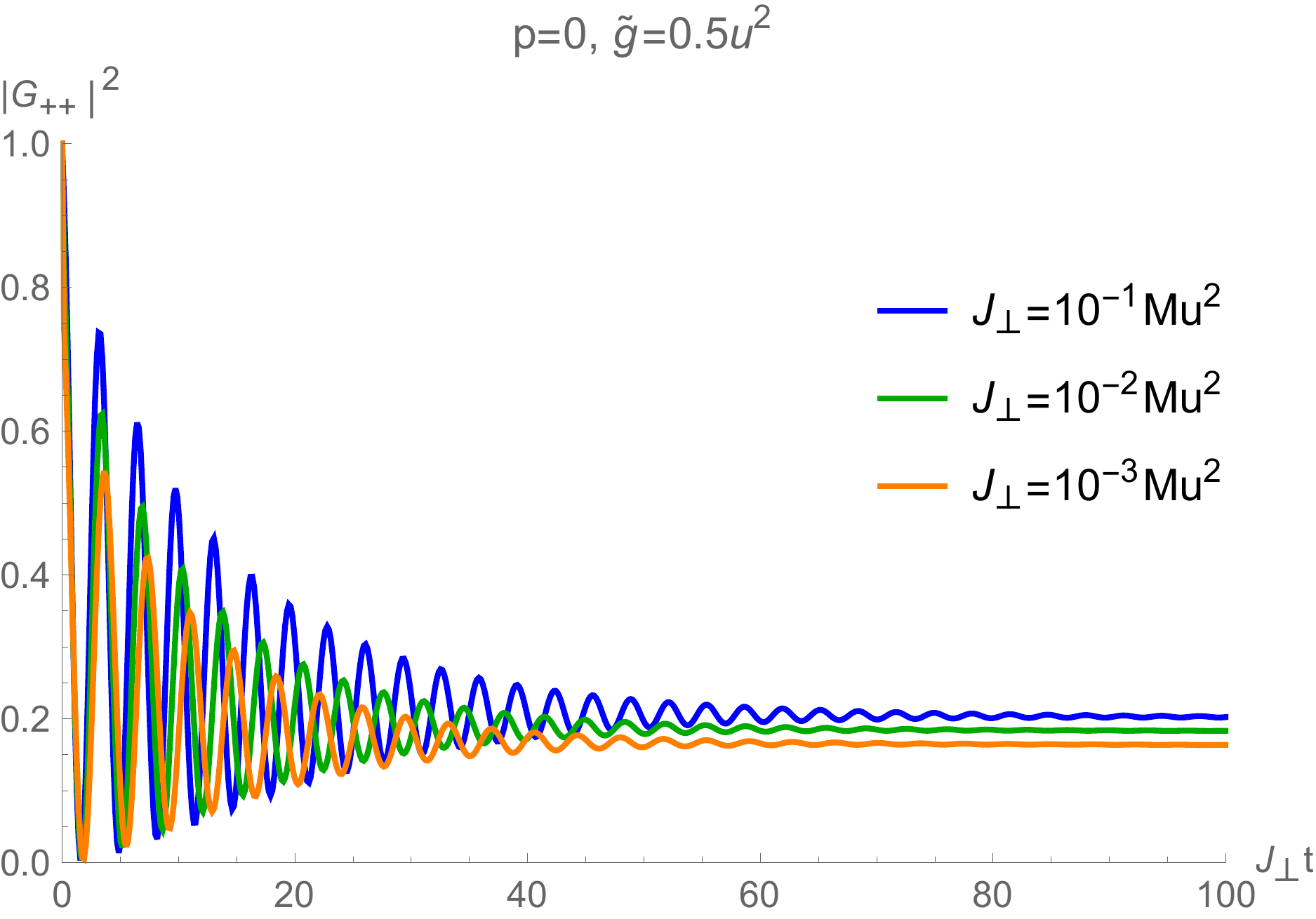}}
    \caption{(Color online) Green's function for symmetric baths, obtained numerically. (a) Coherence is lost at higher $p$. The inset shows that $G_{++}$ and $G_{+-}$ have similar shapes while being out of phase. (b) Increasing $\jp$ causes the oscillations to live relatively longer (please notice that the time scale is $\jp t$). These plots allow to observe the relation between the oscillating and power-law regimes.}
    \label{fig:Gsym1}
\end{figure}
In this \sect~we complement the analytical results reported in the previous one by showing a few plots of the full Green's function, obtained at any time by direct numerical integration of \eq~\eqref{eq:def F and H} (after the simplifications explained in the \app).
\par
\begin{figure}
    \centering
    \includegraphics[width=.9\linewidth]{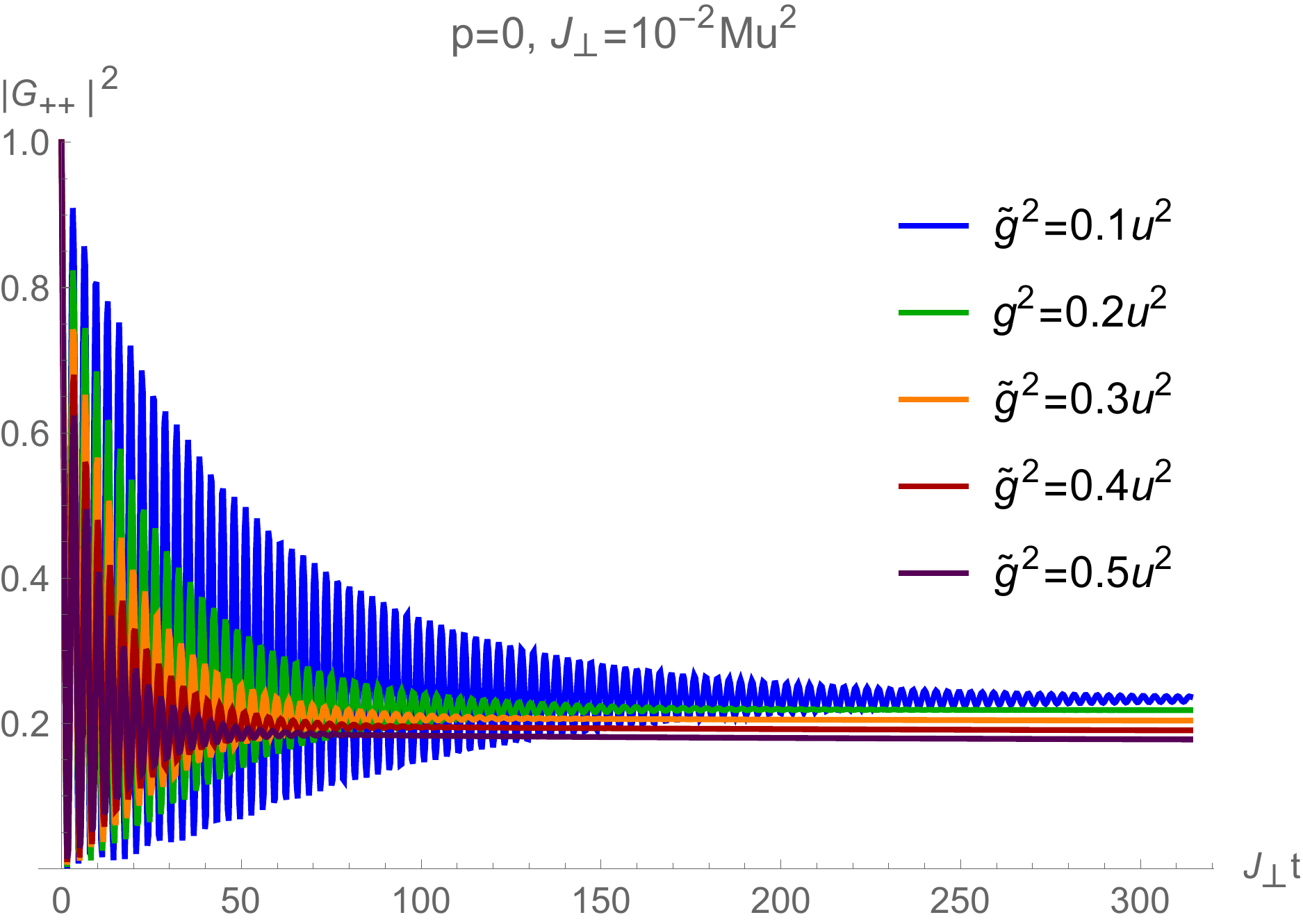}
    \caption{(Color online) Green's function for increasing strength of the impurity bath interaction, showing how larger couplings quench the oscillations and decrease $\abs{G}^2$.}
    \label{fig:Gsym many g}
\end{figure}

 First of all, the simplest case of symmetric baths is shown in \figs~\ref{fig:Gsym1} and \ref{fig:Gsym many g}. The general appearance of the Green's function is the following: $\abs{G_{\sigma\sigma^\prime}}^2$ starts oscillating (at a frequency $2\Tilde{J}_\perp(p)$), with an amplitude that decays as $\ee^{-2\gamma(p)t}$. After a few periods, the oscillations essentially disappear, and the absolute value of both $G_{\parallel}\equiv G_{\sigma\sigma}$ and of $G_\perp\equiv G_{\sigma\bar{\sigma}}$ settle to the same function, which decays as the very weak power law $t^{-2\beta(p)}$. Both the diagonal $G_{++}=G_{--}$ and off-diagonal components have the same overall shape, the only difference being the fact that their oscillations are out of phase (just as in the noninteracting expression, \eq~\eqref{eq: G0}). This is shown in the inset of \fig~\ref{fig:Gsym p}. 
\par All these features are perfectly accounted for by \eq~\eqref{eq: G asymp}, which in fact gives an excellent approximation for all times except for $\jp t\alt 0.1$ when $p$ is small. The oscillations are simply the result of interference of the two terms of \eq~\eqref{eq: G asymp}, and thus they disappear after the component along the antisymmetric mode has decayed (a manifestation of decoherence).
\par \figs~\ref{fig:Gsym p}, \ref{fig:Gsym J} and \ref{fig:Gsym many g} show how changing parameters and momentum alter the Green's function quantitatively. In general, the number of oscillations before the power-law regime decreases as $p$ or $g_\sigma$ increase, or as $\jp$ tends to $0$ (for $\jp=0$ there cannot be any oscillation, obviously). This is consistent with the fact that the decay constant of the antisymmetric mode, $2\gamma(p)$, is an increasing function of $p$, $g_\sigma$ and $\jp$ (see its expression in \eq~\eqref{eq:gamma}). At the same time, it can be noticed that while the number of oscillations decreases also the overall value of the Green's function gets suppressed. This is partially caused by the increase of $\beta(p)$ for larger momentum and/or coupling.
\begin{figure}
    \centering
    \subfloat[][
    \label{fig:Gasym p}]{\includegraphics[width=.9\linewidth]{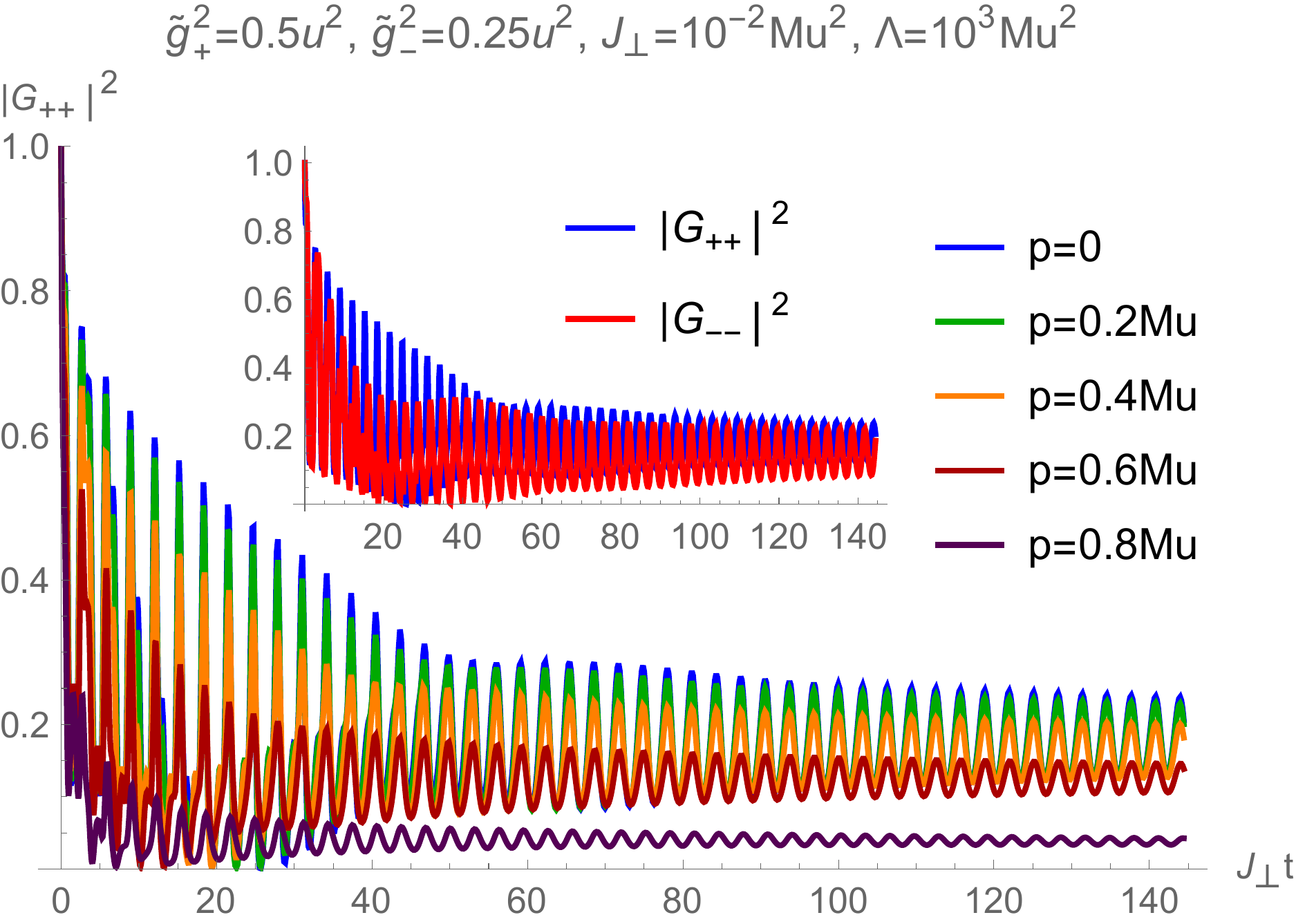}}\\
    \subfloat[][
    \label{fig:Gasym g} ]{\includegraphics[width=.9\linewidth]{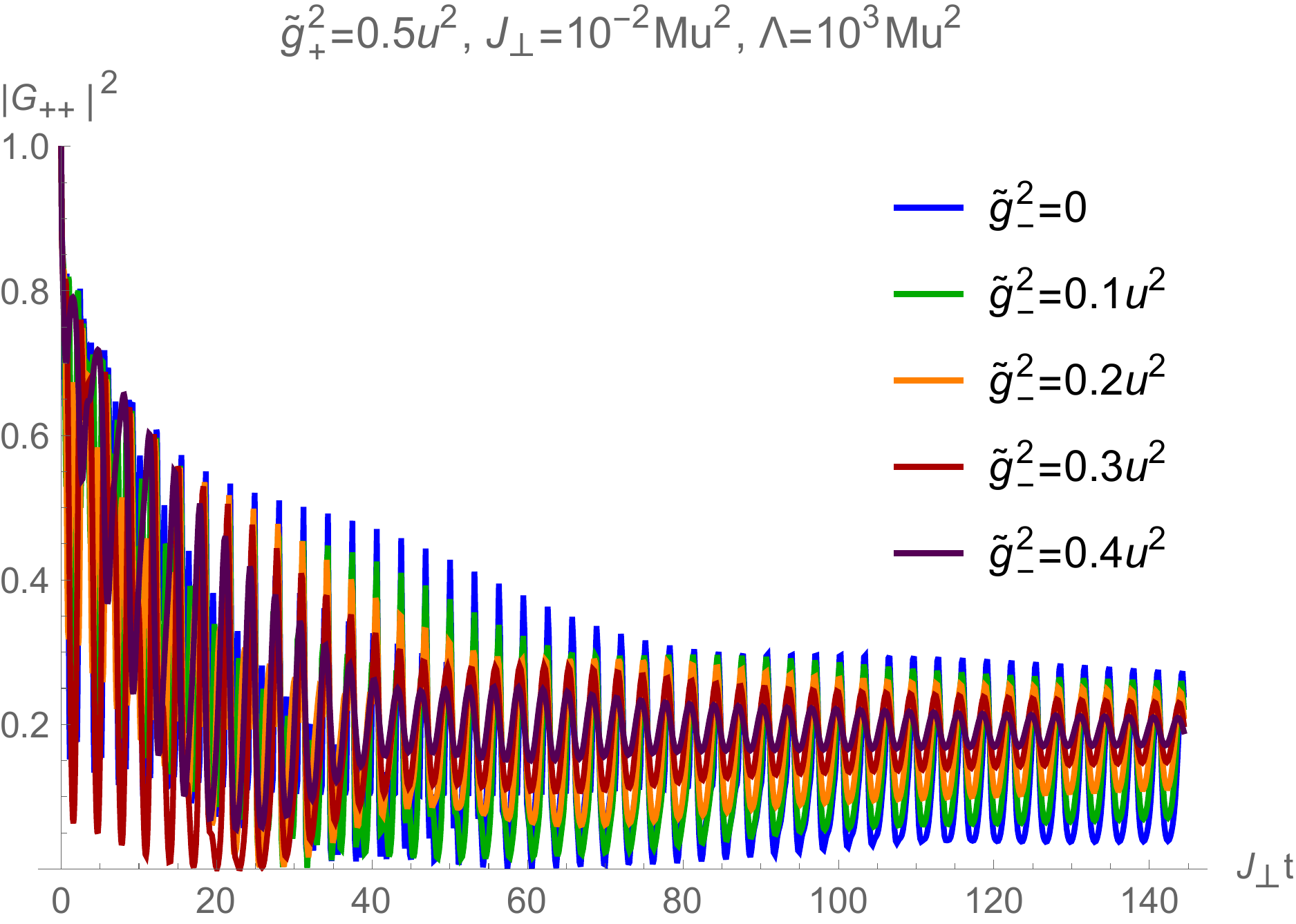}}
    \caption{(Color online) Numerically-obtained $\abs{G_{++}(p,t)}^2$ when $\Tilde{g}_-/u$ is lower than the fixed $\Tilde{g}_+/u$. (a) The Green's function decreases faster at higher momenta. Inset: $\abs{G_{\sigma\sigma}}^2$ is larger on the more interacting bath, although the difference decreases with time. (b) The amplitude of the oscillations increases the more different the baths are.}
    \label{fig:Gasym1}
\end{figure}
\begin{figure}
    \centering
    \includegraphics[width=.9\linewidth]{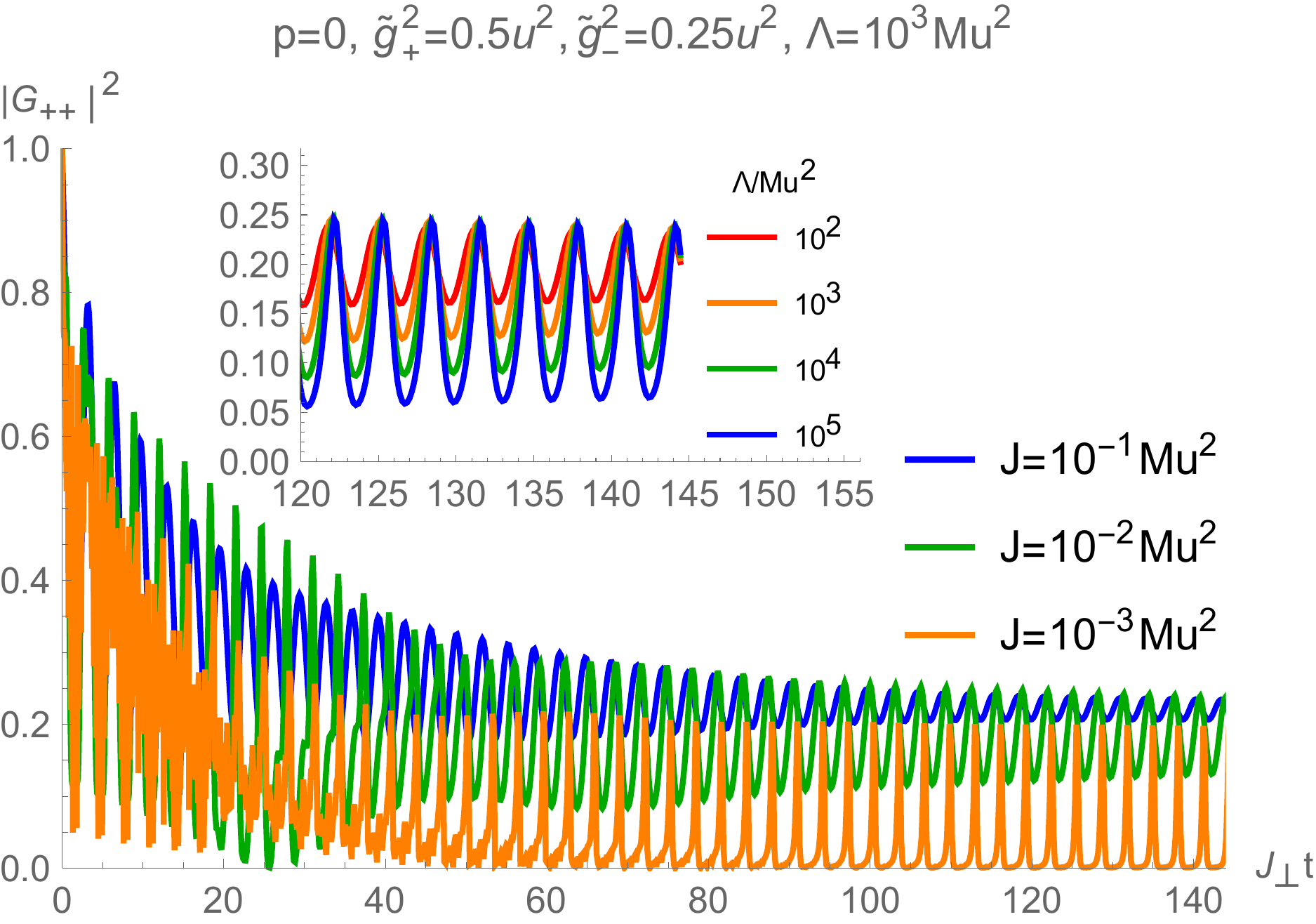}
    \caption{(Color online) Plots of the Green's function for asymmetric baths as the inter-chain hopping is lowered. A smaller $\jp$ causes $\abs{G}^2$ to become noisier at early times and characteristically ''spiked'' at later times. The inset shows that a larger cutoff favours wider oscillations at late times.}
    \label{fig:Gasym J}
\end{figure}
\par When we break the symmetry between the baths the situation changes rather drastically. The asymmetry can be caused by either different bath parameters $u_\sigma,\,K_\sigma$ or by different couplings $g_\sigma$\footnote{Strictly speaking, one could allow also for different cutoffs $\Lambda_\sigma$. We preferred to avoid this complication.}. In our low-momentum description, $g_\sigma$ and $K_\sigma$ get merged into an effective coupling $\tilde{g}_\sigma$, but the actual dimensionless coupling constants are $\tilde{g}_\sigma/u_\sigma$. Therefore, we choose to vary only $\tilde{g}_\sigma$ while keeping the sound speeds constant, but the following qualitative remarks are equally valid if $u_+\neq u_-$.
\par Examples of the Green's function are shown in \figs~\ref{fig:Gasym1} and \ref{fig:Gasym J}. A comparison with the corresponding plots in the symmetric case shows that the overall pattern of increasing or decreasing $\abs{G}^2$ when the parameters are varied is essentially the same. But apart from these large-scale behaviours, the plots are sharply different from the symmetric case. Most prominently, these \figs~generally show wider and more persisting oscillations.
\par Again, there are two different regimes, short times and longer times. The first few oscillation cycles are visibly noisy, with irregular peaks and valleys becoming whose shapes become increasingly irregular for higher momenta and (overall) couplings (\figs~\ref{fig:Gasym p} and \ref{fig:Gasym g}), and especially for the lowest $\jp$s (\fig~\ref{fig:Gasym J}). We have verified that the detailed behaviour in this region is strongly cutoff-dependent. This suggests that it is dominated by the interference between the various terms of \eq~\eqref{eq: full G}, each one depending quite sensitively from the value of $\Lambda$ through the $C(p,t)$ and $D(p,t)$ functions.
\par As time goes on the oscillations acquire a regular shape, with a frequency $2\tilde{J}_\perp(p)$, and a slowly decreasing amplitude. Unless we go to extremely long times, the power-law decline is explicitly seen only by looking at the average. This behaviour does not appear to be related to the variation in $\gamma(p)$ as $\Tilde{g}_\sigma$ are changed. In fact, if we vary the couplings in a way to keep $\gamma(p)$ fixed we get the same results as \fig~\ref{fig:Gasym g}. Moreover, this phenomenon is also rather sharp: a few percent difference between $\Tilde{g}_+$ and $\Tilde{g}_-$ is sufficient to observe it. Once again, its root is the UV logarithmic divergence of the $C(p,t)$ and $D(p,t)$ functions, as confirmed by varying $\Lambda$ - the inset in \fig~\ref{fig:Gasym J} shows how a larger cutoff increases the depth of the oscillations (but leaves the maxima unchanged). 
\par The above discussion shows that even for the large times shown the $C(p,t)$ and $D(p,t)$ functions have still a relevant influence on the Green's function, despite being asymptotically sub-leading. This is particularly true when $\jp$ is very small, as in \fig~\ref{fig:Gasym J}. In this interesting regime, the long-term oscillations have a distinctive ''spiked'', which is very far from all the other cases discussed so far. We can say that thanks to the cutoff-dependence, the coherence between the symmetric and antisymmetric modes is retained longer as soon as the two baths are made unequal.
\subsection{The spectral function}\label{subsect:spectral function}
As it is well known, the spectral function
\begin{equation}
    \hat{A}(p,\omega)\equiv-2\Im\hat{G}(p,\omega)
\end{equation}
yields information about the energy spectrum of the theory\cite{Mahan}. Moreover, we recall that it is also measurable in radio-frequency spectroscpy\cite{PhysRevA.89.053617}. In our case, we can have an insight on what $\hat{A}(p,\omega)$ looks like from the asymptotic expansion, \eq~\eqref{eq: G asymp}.
\par Using the fact that $\hat{G}(p,t)\propto\theta(t)$, we have
\begin{multline*}
    \hat{G}(p,\omega)=\int_{0}^{\infty}{\dd{t}\ee^{\ii\omega^+ t}\hat{G}(p,t)}=\\
    =\underbrace{\int_{0}^{\bar{t}}{\dd{t}\ee^{\ii\omega^+ t}\hat{G}(p,t)}}_{\hat{G}^\textup{reg}}+\underbrace{\int_{\bar{t}}^{\infty}{\dd{t}\ee^{\ii\omega^+ t}\hat{G}(p,t)}}_{\hat{G}^\textup{as}},
\end{multline*}
where $\omega^+=\omega+\ii0^+$ and  $\bar{t}$ is an arbitrary time that is chosen large enough that for later times the Green's function is well approximated by the asymptotic expressions \eqref{eq: G asymp}. Substituting these in $\hat{G}^\textup{as}$ we are left with integrals of the form
\begin{equation*}
    \int_{\bar{t}}^{\infty}{\dd{t}\frac{\ee^{\ii z t}}{t^\beta}}=\frac{\Gamma(1-\beta,-\ii z\bar{t})}{(-\ii z)^{1-
    \beta}}~,
\end{equation*}
where
\begin{equation}
    \Gamma(a,z)\equiv\int_z^{+\infty}{\dd{t}t^{a-1}\ee^{-t}}
\end{equation}
is the incomplete Gamma function\cite{DLMF} (notice that we are interested in weak coupling and small momenta, a regime in which $\beta(p)<1$). Putting them together, we find
\begin{multline*}
    G^\textup{as}_{\sigma\sigma\prime}(p,\omega)\approx-\frac{\ii}{2}t_0^{\beta_p}\Bigg[Z_e(p)\frac{\Gamma(1-\beta_p,-\ii(\omega^+-\Tilde{\lambda}_{pe})\bar{t})}{(-\ii(\omega^+-\Tilde{\lambda}_{pe}))^{1-\beta_p}}+\\
    +\sigma\cdot\sigma^\prime\,Z_o(p)\frac{\Gamma(1-\beta_p,-\ii(\omega^+-\Tilde{\lambda}_{po}+2\ii\gamma_p)\bar{t})}{(-\ii(\omega^+-\Tilde{\lambda}_{po}+2\ii\gamma_p))^{1-\beta_p}}\Bigg]~,
\end{multline*}
where most of the momentum arguments have been transformed into subscripts to improve readability.
\par As it could have been guessed by simple scaling arguments, the power-law decay at long times corresponds to a power-law \emph{divergence} at the frequencies of the renormalized bands\footnote{$\Gamma(a,z)=\Gamma(a)+\order{z}$ for $z\to0$, so it does not change the singular behaviour as long as $a\neq1$, i.e. $\beta(p)\neq1$. For $\beta(p)=1$ one has a logarithmic divergence.}. On the contrary, $\hat{G}^\textup{reg}$ is an integral of a regular function over a finite domain, hence it gives a nonsingular contribution to the Green's function. Therefore, we can conclude that for frequencies around $\Tilde{\lambda}_e(p)$ (and, possibly, also around $\Tilde{\lambda}_o(p)$), $\hat{G}^\textup{as}$ gives the dominant contribution:
\begin{multline}
G_{\sigma\sigma\prime}(p,\omega)\approx-\frac{\ii}{2}t_0^{\beta_p}\Bigg[Z_e(p)\frac{\Gamma(1-\beta_p,-\ii(\omega^+-\Tilde{\lambda}_{pe})\bar{t})}{(-\ii(\omega^+-\Tilde{\lambda}_{pe}))^{1-\beta_p}}+\\
    +\sigma\cdot\sigma^\prime\,Z_o(p)\frac{\Gamma(1-\beta_p,-\ii(\omega^+-\Tilde{\lambda}_{po}+2\ii\gamma_p)\bar{t})}{(-\ii(\omega^+-\Tilde{\lambda}_{po}+2\ii\gamma_p))^{1-\beta_p}}\Bigg]+\\
    +\text{ regular terms for } \omega\to\Tilde{\lambda}_{e,o}
\end{multline}
\begin{figure}
    \centering
    \includegraphics[width=\linewidth]{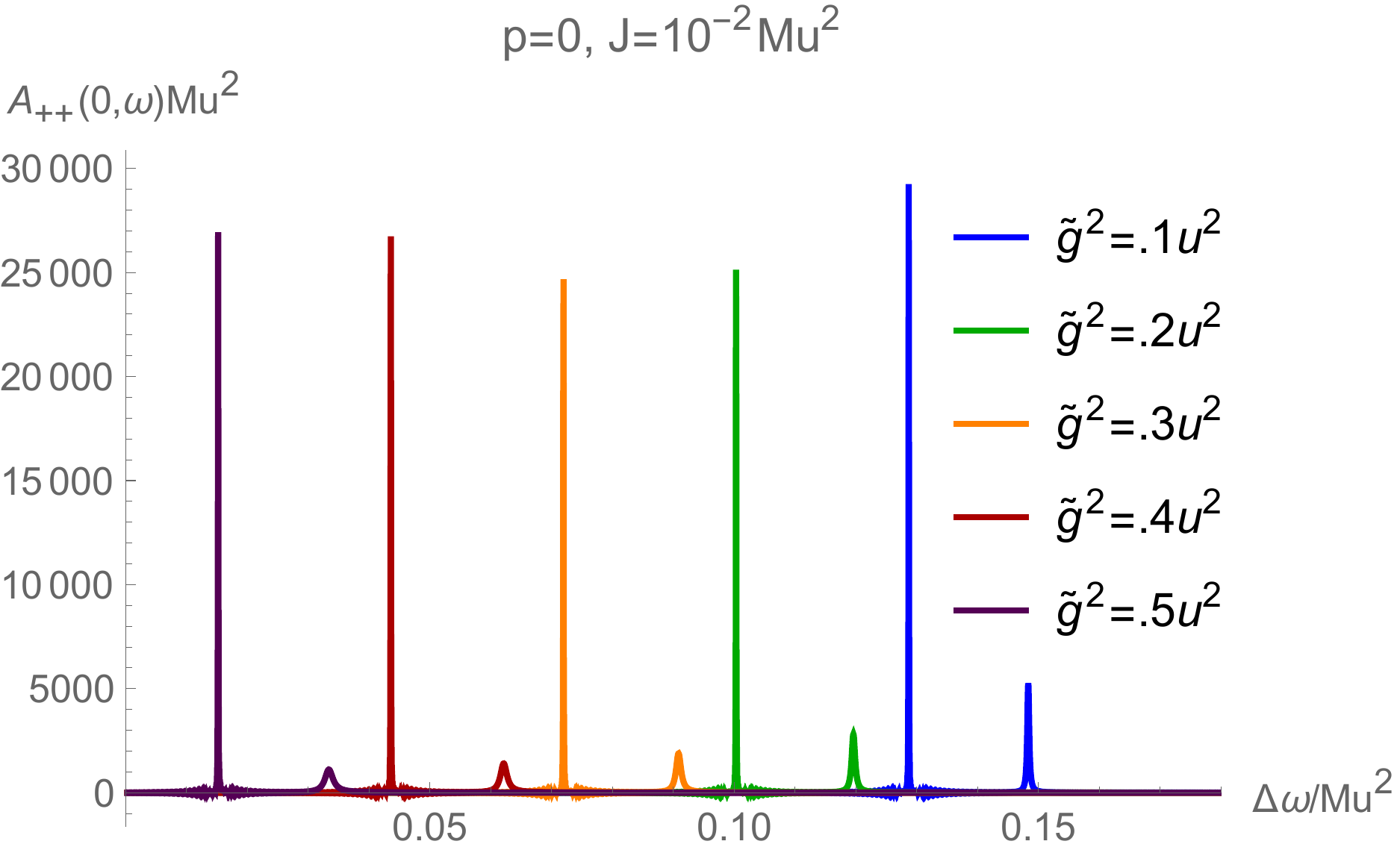}
    \caption{(Color online) Numerically-computed spectral function in the symmetric case. At each coupling the double-peak structure can be appreciated. The oscillations around the higher peaks are numerical artefacts. }
    \label{fig: spectral function}
\end{figure}
When the imaginary part is taken, one observes a double-peak structure, with a threshold-like sharper peak at $\omega=\Tilde{\lambda}_e(p)$ and a broadened one around $\omega=\Tilde{\lambda}_e(p)$. These are, of course, the remnants of the original noninteracting bands.  This can be appreciated clearly in \fig~\ref{fig: spectral function}, which shows the result of the direct numerical computation of $A_{++}(p,\omega)$. 
\par The higher-energy peak corresponds to an unstable state and has a width of order $2\gamma(p)$, but the shape is only approximately Lorentzian. The lower energy one signals a threshold: there are no states with energies below $\Tilde{\lambda}_e(p)$, while immediately above it there is an edge singularity. This can be interpreted as the impurity becoming "dressed" with an arbitrary number of very low-energy phonons. This edge singularity is related to the well-known X-ray threshold problem, and is an established feature of impurities in 1D\cite{PhysRevB.79.241105,Mahan,KSG}. Analytically, it is possible to obtain
\begin{equation}
    A_{\sigma\sigma}(p,\omega)\sim  \theta(\omega-\Tilde{\lambda}_e)t_0^\beta \abs{Z_e(p)}\frac{\Gamma(1-\beta)\sin{\beta\pi}}{(\omega-\Tilde{\lambda}_e(p))^{1-\beta}}\\
\end{equation}
for $\omega\to\Tilde{\lambda}_e^+$. This expression is remarkably similar to the spectral function of the X-ray edge problem\cite{Mahan}.
\subsection{Recovering the single-bath physics}\label{subsect:single bath}
In this last Subsection, we make contact with the results of \rrefs~\onlinecite{KSG} and \onlinecite{PhysRevA.100.023614} for the single-bath setting. We notice that the above-mentioned works focused only on the determination of $\beta^\textup{sb}_\sigma(p)$, and all the other coefficients that we have obtained analytically for the asymptotic behaviour have not been addressed. 
\par The one-bath model is naturally recovered for $\jp\to0$. However, asymptotic expressions such as \eq~\eqref{eq: asympt ABCD} are valid for $t\gg\jp^{-1}$, which is meaningful only if $\jp\neq0$. Therefore, one should start with \eq~\eqref{eq:ABCD} or simply from the definition \eq~\eqref{eq: def F2}. We obtain $\hat{G}(p,t)=\mathrm{diag}(G_+(p,t),G_-(p,t))$, where
\begin{equation}
    G_\sigma(p,t)=-\ii\theta(t)\ee^{-\ii E(p)t+4F_\sigma(\J=0,t)}~,
\end{equation}
in agreement with the calculation of \rrefs\cite{KSG,PhysRevA.100.023614}. Moreover, we have an asymptotic approximation for $F_\sigma(0,t)$ (see \eq~\eqref{eq: asymp F0}), which provides us with the detailed expansion
\begin{multline}
    G_\sigma(p,t)\sim\\
    -\ii Z_\sigma(p, t_0)\ee^{\ii\varphi_\sigma(p)}\qty(\tfrac{t_0}{t})^{\beta^\textup{sb}_\sigma(p)}\ee^{-\ii\Tilde{E}_\sigma(p)t}\big(1+\order{\tfrac{1}{t}}\big)~,
\end{multline}
where
\begin{subequations}
\begin{align}
    Z_\sigma(p,t_0)&=\exp(-\Tilde{g}_\sigma^2\frac{M}{(2\pi)^2}\sum_{s=\pm1}{\frac{\ln{(k_{s\sigma}t_0\ee^{2a_1})}}{2k_{s\sigma}}})~,\\
    \varphi_\sigma(p)&=-\frac{\pi}{2}\beta^\textup{sb}_\sigma(p)~,\\
    \beta^\textup{sb}_\sigma(p)&=\frac{1}{2\pi^2}\frac{g_\sigma^2K_\sigma}{u_\sigma^2}\frac{1+(p/Mu_\sigma)^2}{(1-(p/Mu_\sigma)^2)^2}~,\\
    \Tilde{E}_\sigma(p)&=\frac{p^2}{2M}+\Tilde{g}_\sigma^2\frac{M}{2\pi^2}\ln(\frac{2\abs{p^2-M^2 u_\sigma^2}}{M\Lambda\ee^{-\gamma}})~. 
\end{align}
\end{subequations}
The value of the exponent $\beta^\textup{sb}_\sigma(p)$ that we found agrees with the result reported in \rrefs~\onlinecite{KSG} and \onlinecite{PhysRevA.100.023614}, when expanded to second order in $p/Mu_\sigma$ (except for a difference of a factor $2$ in the first momentum correction for  \rref.~\onlinecite{KSG}).
\section{Conclusions}\label{sec:conclusions}
In this work we have studied the dynamics of an impurity moving in a two-wire interacting system, a system which can be directly realised in the context of ultracold atoms, and it can be considered as a minimal description of a building block of a correlated heterostructure.

Building on the results for an impurity in a single wire, \rref~\onlinecite{KSG}, we have computed the impurity Green's function $\hat{G}$ using the second-order Linked Cluster Expansion. Our results are non-perturbative in the inter-wire hopping. 
\par We have provided detailed analytical expressions for the leading asymptotic terms of $\hat{G}$ in the long-time limit, complementing also the results of  \rref~\onlinecite{KSG} which are mainly numerical.
This allowed us to obtain the renormalisation of the dispersion inside the chain and of the  inter-bath hopping, as well as the exponent of the power-law decay and the width of the antisymmetric mode. 
\par One of our main results is that the orthogonality catastrophe, leading to the breakdown of the quasiparticle picture, survives the inclusion of a second 1D bath and dominates the long-time behaviour of all the components of the Green's function. 

In particular, the exponent characterising the long-time behaviour of the Green's function is given by half of the average of the exponents of the individual baths and, notably, is the same for the intra-wire Green's functions and for that connecting the two wires, demonstrating that, at long times, the behaviour of the system is dominated by the interactions within each wire. This implies that, for this system, the motion inside each wire and the inter-wire motion can not be decoupled, suggesting that extended systems with more chains are necessary for a proper description of the transverse (with respect to the wires) motion of the impurity. 
\par The asymptotic expansion turns out to be very accurate in comparison with a numerical evaluation of the Green's function at arbitrary times. In the case of two nonequivalent baths, the Green's function is non-universal, acquiring a high-frequency component at short times and exhibiting persistent oscillations at longer times. Only at asymptotically large times the symmetric Green's function is recovered. This result may become relevant to the transport properties of heterostructures, which are built out of interfaces between layers of different nature, suggesting that on intermediate timescales we can observe deviation from the behaviour of a single wire. Also in this case, all the components of $\hat{G}$ become asymptotically identical regardless of the inter-wire and intra-wire character. This shows a complete decoherence between the symmetric and antisymmetric states that constitute the noninteracting eigenstates.
\par These results can be directly verified in the context of cold-atom quantum simulators, where the Green's functions can be directly accessed, and used to build an analytical insight on more realistic models for heterostructures featuring layers instead of wires and a larger number of wires/layers.
\par Future work will be devoted to gain a more complete understanding of the dynamics of the system, beyond the partial information provided by the Green's function. An appealing goal would be to address the time evolution of the observables of the system and of the baths, in order to support and complement the conclusions reached in the present work. Moreover, especially in connection with the problem of coherence in heterostructures, it will be stimulating to investigate other possible parameter regimes, such as strong coupling, and settings with more than two baths.
\section{Acknowledgements}
We acknowledge useful discussions with C. Giannetti. This work is supported by Italian MIUR through the PRIN2017 project CEnTral (Protocol Number 20172H2SC4).

\appendix*
\section{Details of the calculations}\label{app: calculations}
\subsection{Expressions of $A,\,B,\,C,\,D$}
The four components of $\hat{F}_2(p,t)$ have the expressions
\begin{subequations}\label{eq:ABCD}
\begin{align}
    A(p,t)&=F(0,t)+\frac{1}{2}\qty(F(\jp,t)+F(-\jp,t))~,\\
    B(p,t)&=\frac{1}{2}\qty(F(-\jp,t)-F(\jp,t))~,\\
    C(p,t)&=\ii\frac{1-\cos{2\jp t}}{\jp}H(0,t)+\nonumber\\&\frac{\sin{\jp t}}{\jp}\qty[-\ee^{-\ii\jp t}H(\jp,t)+\ee^{\ii\jp t}H(-\jp,t)]~,\\
    D(p,t)&=-\ii\frac{\sin{2\jp t}}{\jp}H(0,t)+\nonumber\\
    &\frac{\cos{\jp t}}{\jp}\qty[\ee^{-\ii\jp t}H(\jp,t)-\ee^{\ii\jp t}H(-\jp,t)]~,
\end{align}
\end{subequations}
in terms of the functions
\begin{subequations}
\begin{align}
    F(\J,t)&\equiv\sum_\sigma{F_\sigma(\J,t)}~,\\
    H(\J,t)&\equiv\sum_\sigma{H_\sigma(\J,t)}~,
\end{align}
\end{subequations}
defined as
\begin{subequations}\label{eq:def F and H}
\begin{align}
    F_\sigma(\J,t)&\equiv-\frac{1}{4}\int_{\mathbb{R}}\dd{\ce}\frac{1-\ii \ce t-\ee^{-\ii \ce t}}{\ce^2}R_\sigma(\ce+2\J)~,\\
    H_\sigma(\J,t)&\equiv\sigma\frac{1}{8}\int_{\mathbb{R}}\dd{\ce}\frac{1-\ee^{-\ii \ce t}}{\ce}R_\sigma(\ce+2\J)\nonumber\\
    &\equiv\ii\frac{\sigma}{2}\pdv{F_\sigma(\J,t)}{t}~,
\end{align}
\end{subequations}
for $\J=\pm\jp$ or $0$. The function $R_\sigma(\ce)$ is defined as:
\begin{equation}\label{eq: def R}
    R_\sigma(\ce)\equiv\Tilde{g}_\sigma^2\int_{\mathbb{R}}\tfrac{\dd{q}}{2\pi}V^2(q)\delta(E(p-q)+u_\sigma\abs{q}-E(p)-\ce)~.
\end{equation}
It can be interpreted as the density of states available for scattering between the impurity and the phonons. Expressions like \eq~\eqref{eq:def F and H} and \eq~\eqref{eq: def R} are recurrent when dealing with the OC\cite{SchotteAndSchotte, Mahan}.
\par Recalling that the noninteracting impurity dispersion is given by two bands $\lambda_{e,o}=E(p)\mp\jp$, it can be seen that for $\ce\to0$ $R_\sigma(\ce)$ depends on intra-band processes, while $R_\sigma(\ce\pm2\jp)$ give the effect of inter-band transitions. 
\par Notice that all the results above hold for a generic bare impurity dispersion $E(p)$, as long as it is independent of the bath index.
\par Now we want to take advantage of the quadratic dispersion $E(p)=p^2/2M$, which allows us to explicitly compute $R_\sigma(\ce)$. In the subsonic regime $\abs{p}<M\max_\sigma\{u_\sigma\}$ it reads:
\begin{multline}\label{eq: R expression}
    R_\sigma(\ce)=\Tilde{g}_\sigma^2\frac{M}{(2\pi)^2}\times\\
    \bigg[2-\sum_{s=\pm}{\frac{1}{\sqrt{1+\ce/k_{s\sigma}}}}\bigg]\ee^{-\abs{\ce}/\Lambda}\theta(\ce)~,
\end{multline}
where
\begin{equation}
    k_{s\sigma}\equiv\frac{(Mu_\sigma+sp)^2}{2M},\: s=\pm1~.
\end{equation}
In obtaining \eq~\eqref{eq: R expression}, we have traded the momentum cutoff $\alpha^{-1}$ with an energy cutoff $\Lambda\sim 1/M\alpha^2$, which is easier to handle analytically. The low-energy physics should not be sensitive to the cutoff scheme used.
\subsection{$F_\sigma$ functions}
We have to compute 
\begin{multline}\label{eq: full F}
    F_\sigma(\J,t)=-\frac{\Tilde{g}_\sigma^2}{(4\pi)^2}\int^{\infty}_{-2\J}\dd{\ce}\frac{1-\ii \ce t-\ee^{-\ii \ce t}}{\ce^2}\ee^{-\abs{\ce}/\Lambda}\times\\
    \bigg[2-\sum_{s=\pm}{\frac{1}{\sqrt{1+(2\J+\ce)/k_{s\sigma}}}}\bigg]~,
\end{multline}
Now we manipulate this expression to put it in a form in which it is easier to evaluate numerically, and also to extract its leading  asymptotic terms at large times.
\subsubsection{Case $\J\neq0$}
The first part of the integral in \eq~\eqref{eq: full F} can be calculated exactly: 
\begin{multline}\label{eq:expression f}
    f(\J,t)\equiv2\int^{\infty}_{-2\J}\dd{\ce}\frac{1-\ii \ce t-\ee^{-\ii \ce t}}{\ce^2}\ee^{-\abs{\ce}/\Lambda}=\\
    =\pi t+2t \mathrm{Si}(2\J t)-\frac{1-\cos{2\J t}}{\J}+\\
    2\ii t\bigg[\ln{\tfrac{2\abs{\J}}{\Lambda \ee^{-\gamma}}}+\Re E_1(2\ii\abs{\J} t)+\frac{\sin{2\J t}}{2\J t}\bigg]~,
\end{multline}
where
\begin{equation}
    \mathrm{Si}(z)\equiv\int_0^z\dd{t}\frac{\sin{t}}{t}
\end{equation}
is the sine integral function, and
\begin{equation}
    E_1(z)\equiv\int_z^\infty\dd{t}\frac{\ee^{-t}}{t}
\end{equation}
is the exponential integral function\cite{DLMF}. The symbol $\gamma$ denotes the Euler-Mascheroni constant. Notice that, in the above expression \eqref{eq:expression f} the limit $\Lambda\to\infty$ has been performed whenever this did not cause divergences. As it can be seen, the only part which is dependent on $\Lambda$ is the imaginary part of $f$.
\par When $\jp t\gg1$\footnote{Usually, we expect $\jp\ll k\sim Mu_\sigma^2/2$, i.e. the inter-bath tunnelling amplitude to be far smaller than the intra-bath one.} \eq~\eqref{eq:expression f} has the asymptotic expansion (putting $\J=\pm\jp$)
\begin{subequations}
\begin{align}
    f(\jp,t)&=2\Big(\pi+\ii\ln{\tfrac{2\jp}{\Lambda\ee^{-\gamma}}} \Big)t-\frac{1}{\jp}+\nonumber\\
    +&\ii\frac{\ee^{2\ii\jp t}}{2\jp^2 t}+\order{\frac{1}{(\jp t)^2}}~,\\
    f(-\jp,t)&=2\ii\ln{\tfrac{2\jp}{\Lambda\ee^{-\gamma}}}\,t+\frac{1}{\jp}+\\
    +&\ii\frac{\ee^{-2\ii\jp t}}{2\jp^2 t}+\order{\frac{1}{(\jp t)^2}}~.
\end{align}
\end{subequations}
These expressions have been found using the expansions
\begin{subequations}
\begin{align}
    \mathrm{Si}(z)&\sim \frac{\pi}{2}-\frac{\cos{z}}{z}-\frac{\sin{z}}{z^2}+\order{\frac{1}{z^3}},\quad \abs{z}\gg1\\
    E_1(z)&\sim\frac{\ee^{-z}}{z}\Big(1-\frac{1}{z}+\order{\frac{1}{x^2}}\Big),\quad\abs{z}\gg1
\end{align}
\end{subequations}
and the fact that $\mathrm{Si}(x)$ is odd.
\par The remaining integrals are in the form
\begin{equation*}
    \int^{\infty}_{-2\J}\dd{\ce}\frac{1-\ii\ce t-\ee^{-\ii \ce t}}{\ce^2}\frac{1}{\sqrt{1+(2\J+\ce)/k}}~,
\end{equation*}
and are finite when the cutoff is removed. The above \eq~can be brought to a simpler form by doing the integral in the complex plane. The main goal is to integrate parallel to the imaginary axis, i.e. $\ce\to \ii u$, so that the integral goes from oscillatory to exponentially damped, allowing for a more efficient numerical integration. As a byproduct, some pieces of the resulting expression can be computed analytically. Moreover, this provides a shorter route to the asymptotic form when $t\to\infty$.
\begin{figure}
    \centering
    \includegraphics[width=.86\linewidth]{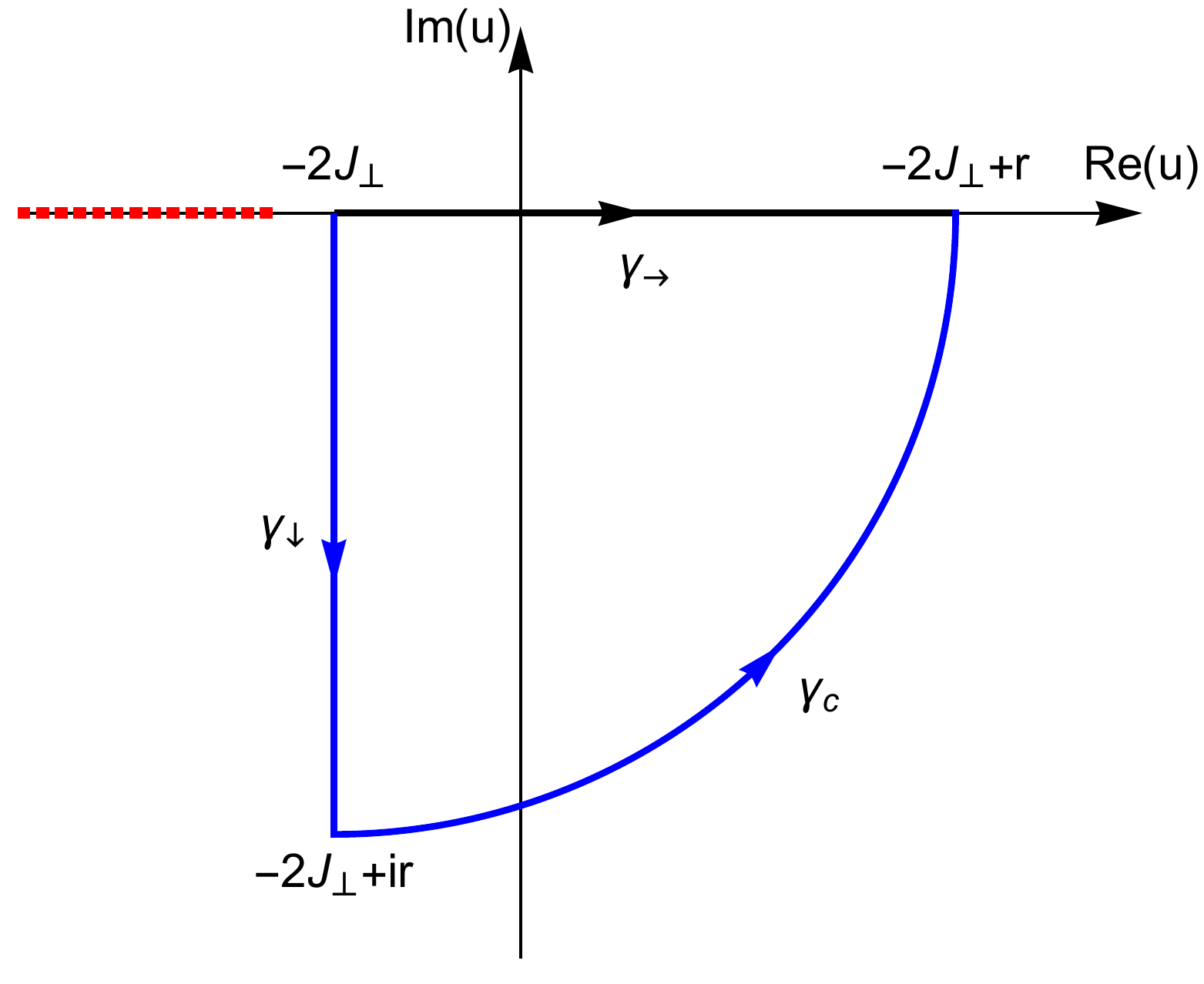}
    \caption{(Color online) Paths of integration in the complex plane. Black: original path. Blue: deformed path. Dotted red: branch cut. It has been chosen to show the $\J>0$ case, the $\J\leq0$ ones are analogous.}
    \label{fig:complex path}
\end{figure}
\par One first deforms the integration path as depicted in \fig~\ref{fig:complex path}, from a segment on the real axis to a vertical segment followed by a quarter of circumference of radius $r$. This does not change the value of the integral, because the only singularity of the integrand is the square root branch cut, which is chosen to lie on the real axis, to the left of $-k-2\J$. Moreover, it is not difficult to see that the integral on the quarter of circumference vanishes as $1/\sqrt{r}$ when $r\to\infty$, and one obtains:
\begin{multline}\label{eq: contour integral}
    \int_{-2\J}^{\infty}\dd{\ce}\frac{1-\ii\ce t-\ee^{-\ii\ce t}}{u^2}\frac{1}{\sqrt{1+(\ce+2\J)/k}}=\\
    =-\ii\int_0^{\infty}\dd{u}\frac{1+\ii (\ii u+2\J)t-\ee^{2\ii\J t- ut}}{(\ii u+2\J)^2}\frac{1}{\sqrt{1-\ii u/k}}=\\
    =-\ii\frac{1}{k}\phi_1(\tfrac{2\J}{k})+t \phi_2(\tfrac{2\J}{k})+\\
    -\ii\ee^{2\ii\J t}\int_0^{\infty}\dd{u}\frac{\ee^{-u t}}{(\ii u+2\J)^2}\frac{1}{\sqrt{1-\ii u/k}}~,
\end{multline}
where
\begin{subequations}\label{eq: def phi}
\begin{align}
    \phi_1(x)&\equiv \int_0^{\infty}\dd{u}\frac{1}{(\ii u+x)^2}\frac{1}{\sqrt{1-\ii u}}\\
    \phi_2(x)&\equiv \int_0^{\infty}\dd{u}\frac{1}{\ii u+x}\frac{1}{\sqrt{1-\ii u}}
\end{align}
\end{subequations}
This is the desired expression.The two functions \eqref{eq: def phi} can be computed exactly by going back to the real axis\footnote{One may exploit the observation that $\phi_1(x)=-\dv{\phi_2}{x}$.}:
\begin{equation}
    \phi_1(x)=
    \begin{cases}
    \frac{1}{2x(1+x)^{3/2}}\Big[\pi x+\nonumber\\
    \;-2\ii\qty(\sqrt{x+1}+x\, \ash(\tfrac{1}{\sqrt{x}}))\Big],\quad x>0\\
    \frac{\ii}{\abs{x}(1-\abs{x})}-\frac{\ii}{(1-\abs{x})^{3/2}}\ach(\tfrac{1}{\sqrt{\abs{x}}}),\:\: x<0
    \end{cases}~,
\end{equation}
\begin{equation}
    \phi_2(x)=
    \begin{cases}
    \frac{1}{\sqrt{1+x}}\Big[\pi-2\ii\,\ash(\tfrac{1}{\sqrt{x}})\Big],\quad x>0\\
    -\frac{2\ii}{\sqrt{1-\abs{x}}}\ach(\tfrac{1}{\sqrt{\abs{x}}}),\quad x<0
    \end{cases}~,
\end{equation}
in which it is understood that
\begin{equation*}
    \ach(x)=\ii\arccos{x}\qq{for}\abs{x}<1~.
\end{equation*}
The remaining integral is fast-converging when performed numerically. 
\par When $\jp t\gg1$ one can easily find an asymptotic approximation of this term, using
\begin{equation}\label{eq:asymp Laplace}
    \int_a^\infty{\dd{u}\ee^{-t u}q(u)}\sim\ee^{-a t}\sum_{n=0}^{\infty}{\frac{q^{(n)}(0)}{t^{n+1}}}~,
\end{equation}
when the function $q(u)$ is infinitely differentiable around $u=a$ (see \cite{DLMF}). This can be applied to the last integral in \eq~\eqref{eq: contour integral}:
\begin{equation*}
    \int_0^{\infty}\dd{u}\frac{\ee^{-u t}}{(\ii u+2\J)^2}\frac{1}{\sqrt{1-\ii u/k}}=\frac{1}{(2\J)^2 t}+\order{\frac{1}{(\J t)^2}}~.
\end{equation*}
Thus, the contribution of this term to the integral becomes small quite rapidly in the long-time limit.
\par To sum up, we can write
\begin{multline}
     \int_{-2\J}^{\infty}\dd{\ce}\frac{1-\ii\ce t-\ee^{-\ii\ce t}}{\ce^2}\frac{1}{\sqrt{1+(\ce+2\J)/k}}=\\=t \phi_2(\tfrac{2\J}{k})-\ii\frac{1}{k}\phi_1(\tfrac{2\J}{k})-\ii\frac{\ee^{2\ii\J t}}{(2\J)^2 t}+\order{\frac{1}{(2\J t)^2}}~,
\end{multline}
and putting it together with \eq~\eqref{eq:expression f}, we find the sought asymptotic approximation:
\begin{subequations}
\begin{align}
    F_\sigma(\jp,t)=-\Tilde{g}_\sigma^2\frac{M}{(4\pi)^2}&\Big[-\frac{1}{\jp}+\ii\sum_{s=\pm1}{\frac{1}{k_{s\sigma}}\phi_1(\tfrac{2\jp}{k_{s\sigma}}})+\nonumber\\
    +\big(2\pi+2\ii&\ln{\tfrac{2\jp}{\Lambda\ee^{-\gamma}}-\sum_{s=\pm1}{\phi_2(\tfrac{2\jp}{k_{s\sigma}})}}\big)t+\nonumber\\    +\ii\frac{\ee^{2\ii\jp t}}{\jp^2 t}&+\order{\frac{1}{(\jp t)^2}}\Big]\\
    F_\sigma(-\jp,t)=-\Tilde{g}_\sigma^2\frac{M}{(4\pi)^2}&\Big[\frac{1}{\jp}+\ii\sum_{s=\pm1}{\frac{1}{k_{s\sigma}}\phi_1(-\tfrac{2\jp}{k_{s\sigma}})}+\nonumber\\
     +\big(2\ii\ln{\tfrac{2\jp}{\Lambda\ee^{-\gamma}}}&-\sum_{s=\pm1}{\phi_2(-\tfrac{2\jp}{k_{s\sigma}})}\big)t+\nonumber\\    +\ii\frac{\ee^{-2\ii\jp t}}{\jp^2 t}&+\order{\frac{1}{(\jp t)^2}}\Big]~.
\end{align}
\end{subequations}
\subsubsection{Case $\J=0$}
Once again, the first part of \eq~\eqref{eq: full F} is calculated exactly:
\begin{equation}
    f(0,t)\equiv2\int^{\infty}_{0}\dd{\ce}\frac{1-\ii\ce t-\ee^{-\ii\ce t}}{\ce^2}\ee^{-\abs{\ce}/\Lambda}=\pi t-2\ii t\ln{\tfrac{\Lambda t}{\ee}}~.
\end{equation}
Already at this stage, a term proportional to $\ln{t}$ has appeared.
\par The remaining integrals, which are of the form
\begin{multline}
    \int_{0}^{\infty}\dd{\ce}\frac{1+\ii\ce t-\ee^{-\ii\ce t}}{\ce^2}\frac{1}{\sqrt{1+\ce/k}}=\\
    \ii\int_0^{\infty}\dd{u}\frac{1-ut-\ee^{-ut}}{ u^2}\frac{1}{\sqrt{1-\ii u/k}}
\end{multline}
cannot be separated into three components as in the $\jp\neq0$ case, because the "kernel" $(1- ut-\ee^{-ut})/u^2$ has to be treated carefully at the $u=0$ integration limit. One can do the following:
\begin{multline}
    \int_{0}^{\infty}\dd{\ce}\frac{1-\ii\ce t-\ee^{-\ii\ce t}}{\ce^2}\frac{1}{\sqrt{1+\ce/k}}=\\
    =t\Big[\underbrace{\int_0^{\bar{\ce}}{\dd{\ce}\frac{1-\ii \ce- \ee^{-\ii \ce}}{\ce^2}\frac{1}{\sqrt{1+\ce/\bar{\ce}}}}}_{I}+\\
    +\underbrace{\int_{\bar{\ce}}^\infty{\dd{\ce}\frac{1-\ii \ce- \ee^{-\ii \ce}}{\ce^2}\frac{1}{\sqrt{1+\ce/\bar{\ce}}}}}_{II}\Big]~,
\end{multline}
with the substitution $\ce\to\ce t$ and using the notation $\bar{\ce}\equiv kt$. The first integral can be estimated for large $t$, i.e. for large $\bar{\ce}$, by expanding the square root in powers of $\ce/\bar{\ce}$:
\begin{multline}
    I\sim\sum_{n=0}^\infty{\binom{-1/2}{n}\frac{1}{\bar{\ce}^n}\int_0^{\bar{\ce}}{\dd{\ce}\frac{1-\ii \ce- \ee^{-\ii \ce}}{\ce^2}}\ce^n}=\\
    =\frac{\pi}{2}-\big(1+\frac{\gamma}{2}-c_1\big)\frac{1}{\bar{\ce}}-\frac{\ln{\bar{\ce}}}{\bar{\ce}}+\\
    +\ii\big(1-\gamma-c_0-\ln{\bar{\ce}}-\frac{\pi}{4\bar{\ce}}\big)+\order{\tfrac{1}{\bar{\ce}^2}}~,
\end{multline}
where
\begin{subequations}
\begin{align}
    c_0&\equiv\sum_{n\ge1}{\binom{-1/2}{n}\frac{1}{n}}=\ln{4}-2\ash(1)~,\\
    c_1&\equiv\sum_{n\ge2}{\binom{-1/2}{n}\frac{1}{n-1}}=\nonumber\\
    &\quad=\frac{3-2\sqrt{2}}{2}-\ln{2}+\ash(1)~.
\end{align}
\end{subequations}
To obtain these results, we integrated explicitly all terms in the series and expanded them up to $\order{\bar{\ce}^{-2}}$. As one can see, terms containing $\ln{t}$ have appeared. These are due to the structure of the kernel $(1+\ii \ce -\ee^{\ii \ce})/\ce^2$ for $\ce\sim0$.
\par The second integral avoids the neighbourhood of $\ce=0$ and therefore it can be computed straightforwardly:
\begin{multline}
    II=\underbrace{\int_{\bar{\ce}}^\infty{\dd{\ce}\frac{1}{\ce^2\sqrt{1+\ce/\bar{\ce}}}}}_{II_a}+\\
    \underbrace{-\ii\int_{\bar{\ce}}^\infty{\dd{\ce}\frac{1}{\ce\sqrt{1+\ce/\bar{\ce}}}}}_{II_b}+\\
    \underbrace{-\int_{\bar{\ce}}^\infty{\dd{\ce}\frac{\ee^{-\ii \ce}}{\ce^2\sqrt{1+\ce/\bar{\ce}}}}}_{II_c}~,
\end{multline}
giving
\begin{gather*}
    II_a=\frac{1}{\bar{\ce}}\int_1^\infty{\dd{\ce}\frac{1}{\ce^2\sqrt{1+\ce}}}=\frac{\sqrt{2}-\ash(1)}{\bar{\ce}}~,\\
    II_b=-\ii\int_1^\infty{\dd{\ce}\frac{1}{\ce\sqrt{1+\ce}}}=-2\ii\ash(1)~,\\
   II_c= -\int_{\bar{\ce}}^\infty{\dd{\ce}\frac{\ee^{-\ii \ce}}{\ce^2\sqrt{1+\ce/\bar{\ce}}}}\sim \frac{\ii\ee^{-\ii \bar{\ce}}}{\sqrt{2}\bar{\ce}^2}+\order{\tfrac{1}{\bar{\ce}^3}},\; \bar{\ce}\gg1~,
\end{gather*}
where for the last estimate we used the analogous of \eq~\eqref{eq:asymp Laplace}\cite{DLMF}:
\begin{equation}\label{eq:asymp Fourier}
    \int_a^{\infty}{\dd{\ce}\ee^{\ii x \ce}q(\ce)}\sim\ee^{\ii a x}\sum_{n=0}^\infty{q^{(n)}(a)\big(\tfrac{\ii}{x}\big)^{n+1}},\quad x\gg1~.
\end{equation}
Hence,
\begin{equation*}
    II\sim -2\ii\ash(1)+\frac{\sqrt{2}-\ash(1)}{\bar{\ce}}+\frac{\ii\ee^{-\ii \bar{\ce}}}{\sqrt{2}\bar{\ce}^2}+\order{\tfrac{1}{\bar{\ce}^3}}~,
\end{equation*}
which, unlike the $I$ integral, does not contain any term depending logarithmically on $\bar{\ce}$.
\par The final estimate is 
\begin{gather}
    \int_{0}^{\infty}\dd{\ce}\frac{1+\ii\ce t-\ee^{\ii\ce t}}{\ce^2}\frac{1}{\sqrt{1+\ce/k}}=\nonumber\\
    =t(I+II)=-\frac{a_1}{k}-\frac{\ln{kt}}{2k}+\frac{\pi}{2}t+\nonumber\\
    +\ii\Big[-\frac{\pi}{4k}-t\ln{kt}+(1-\gamma-\ln{4})t\Big]+\order{\tfrac{1}{t}}~,
\end{gather}
where
\begin{equation}
    a_1\equiv\ln{2}-\frac{1-\gamma}{2}~.
\end{equation}
Putting all terms together, we obtain the asymptotic approximation
\begin{multline}\label{eq: asymp F0}
    F_\sigma(0,t)\sim-\Tilde{g}_\sigma^2\frac{M}{(4\pi)^2}\Big[\big(2a_1+\ii\frac{\pi}{2}\big)\sum_{s=\pm1}{\frac{M}{(p+sMu_\sigma)^2}}+\\
    +\sum_{s=\pm1}{\frac{M}{(p+sMu_\sigma)^2}}\ln{\frac{(p+sMu_\sigma)^2t}{2M}}+\\
    +2\ii t\ln{\frac{2\abs{p^2-M^2u_\sigma^2}}{M\Lambda\ee^{-\gamma}}+\order{\tfrac{2M}{(p\pm M u_\sigma)^2 t}}}\Big],\\ \qq{for}\frac{(p\pm M u_\sigma)^2 t}{2M}\gg1~.
\end{multline}
\subsection{$H_\sigma$ functions}
The $H_\sigma$ functions are obtained from the $F_\sigma$ ones by
\begin{equation*}
    H_\sigma(\J,t)=\ii\frac{\sigma}{2}\pdv{F_\sigma(\J,t)}{t}~.
\end{equation*}
The fast-converging integral expression is
\begin{multline}
    H_\sigma(\J,t)=\sigma\frac{\Tilde{g}_\sigma^2 M}{2(4\pi)^2}\Big[h(\J,t)+\\
    +\ii\sum_{s=\pm1}{\Big(\phi_2(\tfrac{2\J}{k_{s\sigma}})-\int_0^\infty{\dd{\ce}\frac{\ee^{2\ii\J t-t \ce}}{\ii \ce+2\J}\frac{1}{\sqrt{1-\ii \ce/k_{s\sigma}}}}\Big)}\Big]~,
\end{multline}
where
\begin{subequations}
\begin{align}
    h(\pm\jp,t)&=2\ln{\tfrac{2\jp}{\Lambda\ee^{-\gamma}}}+2\Re E_1(2\ii\jp t)+\nonumber\\
    &-\ii\big(\pi\pm2\mathrm{Si}(2\jp t)\big)\sim\nonumber\\
    &\sim2\ln{\tfrac{2\jp}{\Lambda\ee^{-\gamma}}}-\ii(1\pm1)\pi+\nonumber\\
    &\pm\ii\frac{\ee^{\pm2\ii\jp t}}{\jp t}+\order{\tfrac{1}{(\jp t)^2}}\\
    h(0,t)&=-2\ln{\Lambda t}-\ii\pi
\end{align}
\end{subequations}
The asymptotic expansions are
\begin{subequations}
\begin{align}
    H_\sigma(\jp,t)=\sigma\frac{\Tilde{g}_\sigma^2 M}{2(4\pi)^2}&\Big(2\ln{\tfrac{2\jp}{\Lambda\ee^{-\gamma}}}-2\ii\pi+\nonumber\\
    +\ii\sum_{s=\pm1}{\phi_2(\tfrac{2\jp}{k_{s\sigma}})}&-2\ii\frac{\ee^{2\ii\jp t}}{\jp t}+\order{\tfrac{1}{(\jp t)^2}}\Big)\\
    H_\sigma(-\jp,t)=\sigma\frac{\Tilde{g}_\sigma^2 M}{2(4\pi)^2}&\Big(2\ln{\tfrac{2\jp}{\Lambda\ee^{-\gamma}}}+\nonumber\\
    +\ii\sum_{s=\pm1}{\phi_2(-\tfrac{2\jp}{k_{s\sigma}})}&+2\ii\frac{\ee^{-2\ii\jp t}}{\jp t}+\order{\tfrac{1}{(\jp t)^2}}\Big)\\
    H_\sigma(0,t)=\sigma\frac{\Tilde{g}_\sigma^2 M}{2(4\pi)^2}&\Big(2\ln\tfrac{2\abs{p^2-M^2u_\sigma^2}}{M\Lambda\ee^{-\gamma}}+\nonumber\\
    -\ii\sum_{s=\pm1}{\frac{1}{2k_{s\sigma}t}}&+\order{\tfrac{1}{(\jp t)^2}}\Big)
\end{align}
\end{subequations}
As one can see, all these functions tend to a constant at $t\to\infty$ (as they should, being derivatives of asymptotically linear functions $F_\sigma$).
\subsection{Asymptotic expansion of $A,\,B,\,C,\,D$.}
Above, we found the necessary ingredients to compute the leading asymptotic behaviour of the $\hat{F}_2(p,t)$ function. The result is \eq~\eqref{eq: asympt ABCD}, with the coefficients given by
\begin{subequations}\label{eq: AB coefficients}
\begin{align}
    \beta(p)&=\frac{M}{2(4\pi)^2}\sum_{s,\sigma}{\frac{\Tilde{g}_\sigma^2}{k_{s\sigma}}}~,\\
    \gamma(p)&=\frac{M}{32\pi}\sum_{\sigma,s}{\Tilde{g}_\sigma^2\Big(1-\frac{1}{\sqrt{1+2\jp/k_{s\sigma}}}\Big)}~,\\
    c_A(p,t_0)&=-\frac{M}{2(4\pi)^2}\sum_{s,\sigma}{\frac{\Tilde{g}_\sigma^2}{k_{s\sigma}}\Big[2a_1+\ln{k_{s\sigma}t_0}}+\nonumber\\
    &+\ii\qty(\phi_1(\tfrac{2\jp}{k_{s\sigma}})+\phi_1(-\tfrac{2\jp}{k_{s\sigma}})+\tfrac{\pi}{2})\Big]~,\\
    c_B(p)&=-\frac{M}{2(4\pi)^2}\sum_{s,\sigma}{\frac{\Tilde{g}_\sigma^2}{k_{s\sigma}}\Big[\tfrac{k_{s\sigma}}{\jp}}+\nonumber\\
    &+\ii\qty(\phi_1(\tfrac{-2\jp}{k_{s\sigma}})-\phi_1(\tfrac{2\jp}{k_{s\sigma}}))\Big]~,\\
    \Delta E(p)&= \frac{M}{(4\pi)^2}\sum_{s,\sigma}{\Tilde{g}_\sigma^2\bigg[\ln(\tfrac{4\jp}{\Lambda\ee^{-\gamma}}\tfrac{\abs{p^2-M^2u_\sigma^2}}{M\Lambda\ee^{-\gamma}})}+\nonumber\\
    &+\tfrac{\ash(\sqrt{k_{s\sigma}/2\jp})}{\sqrt{1+2\jp/k_{s\sigma}}}+\tfrac{\ach(\sqrt{k_{s\sigma}/2\jp})}{\sqrt{1-2\jp/k_{s\sigma}}}\bigg]~,\\
    \Delta J(p)&=-\frac{M}{(4\pi)^2}\sum_{s,\sigma}{\Tilde{g}_\sigma^2\bigg[\tfrac{\ach(\sqrt{k_{s\sigma}/2\jp})}{\sqrt{1-2\jp/k_{s\sigma}}}}+\nonumber\\
    &-\tfrac{\ash(\sqrt{k_{s\sigma}/2\jp})}{\sqrt{1+2\jp/k_{s\sigma}}}\bigg]~,
\end{align}
\end{subequations}
and
\begin{subequations}\label{eq: CD coefficients}
\begin{align}
    c^{(+)}_H(p)&=\frac{M}{2(4\pi)^2}\sum_{s,\sigma}{\sigma\Tilde{g}_\sigma^2\Big(\ln{\tfrac{2\jp}{\Lambda\ee^{-\gamma}}}-\ii\pi+\ii\phi_2(\tfrac{2\jp}{k_{s\sigma}})\Big)}~,\\
    c^{(-)}_H(p)&=\frac{M}{2(4\pi)^2}\sum_{s,\sigma}{\sigma\Tilde{g}_\sigma^2\Big(\ln{\tfrac{2\jp}{\Lambda\ee^{-\gamma}}}+\ii\phi_2(-\tfrac{2\jp}{k_{s\sigma}})\Big)}~,\\
    c^{(0)}_H(p)&=\frac{M}{(4\pi)^2}\sum_{\sigma}{\sigma \Tilde{g}_\sigma^2\ln\tfrac{2\abs{p^2-M^2u_\sigma^2}}{M\Lambda\ee^{-\gamma}}}~,
\end{align}
\end{subequations}
where $t_0$ is an arbitrary time scale (so that the physical dimensions are consistent).
\par One should notice that only $\Delta E(p)$ and the three $c_H(p)$s bear a (logarithmic) dependence on the cutoff $\Lambda$. The former is an energy shift, so it is reasonable that it depends sensitively on the behaviour of the theory at high energy. Conversely, measurable quantities like $\Tilde{J}_\perp\equiv\jp+\Delta J$  and $\beta(p)$ are cutoff-independent and thus proper low-energy properties.
\par We would also like to do some remarks about the region of validity of the asymptotic expansions presented in this work. We have found that, numerically, the expressions \eq~\eqref{eq: asympt ABCD} give rather accurate results even for times $\jp t\sim .1$, at least for $\abs{\hat{G}(p,t)}$ and $p\alt.5$. Increasing the momentum causes one of the  $k_{s\sigma}(p)$ to become small, and hence the asymptotic expressions retain their reliability only at larger times. This is also valid for the approximation given in \eq~\eqref{eq: G asymp1} or \eqref{eq: G asymp} in the case of symmetric baths. For increasing asymmetry \eqs~\eqref{eq: G asymp1} and \eqref{eq: G asymp} are reliable only for larger times, depending on the parameters used.
\bibliography{biblioPaper.bib}
\end{document}